\documentstyle[eqsecnum,aps,epsf]{revtex}
\newcommand\beq{\begin{equation}}
\newcommand\eeq{\end{equation}}
\begin{document}
\draft 
\title{
\hfill {\normalsize KUNS-1535 }\\
Thermal and Quantum Fluctuations  in the Kaon Condensed Phase}
\author{Toshitaka Tatsumi and Masatomi Yasuhira
\footnote{E-mail address: tatsumi@ruby.scphys.kyoto-u.ac.jp; 
\qquad yasuhira@ruby.scphys.kyoto-u.ac.jp}\\ 
Department of Physics, Kyoto University\\
Kyoto 606-8502, Japan}
\date{\today}
 
\maketitle
 
\begin{abstract}
A new formulation is presented to treat thermal and quantum
fluctuations around 
the kaon condensate on the basis of chiral symmetry; 
separating the zero-mode from the beginning and using the
path-integral method, we
can formulate the inclusion of fluctuations in a transparent
way.  Nucleons as well as kaons are treated in a self-consistent way
to the one-loop order.
The effects of the Goldstone mode, stemming from the breakdown of
$V-$ spin symmetry in the condensed phase, are figured out. 

A procedure is discussed to renormalize the divergent integrals properly 
up to the one-loop order. Consequently the thermodynamic potential is
derived. It is pointed out that the zero-point fluctuation by nucleons gives 
a sizable effect, different from the kaonic one. 
  

\end{abstract}
\section{Introduction}

It has been extensively discussed that kaons, the lightest hadrons
with strangeness, show characteristic
features through the kaon-nucleon interactions once immersed in 
nuclear medium. 
The scalar interaction (called the sigma term) and the vector
interaction (called the Tomozawa-Weinberg term) are the leading ones,
and they bring about a large attraction for the excitation energy of $K^-$ 
mesons at low momentum, whereas a repulsion for $K^+$ mesons. There
are many proposals to see these consequences experimentally; $K^-$
enhancement in the subthreshold kaon production or the anomaly around the
$\phi$ meson peak in the dilepton production from the relativistic
heavy-ion collisions are some of them \cite{kaos,li,fuj}.

In 1986 Kaplan and Nelson suggested a possibility of kaon condensation 
in dense matter by using a chiral Lagrangian \cite{kapl}. 
Since then
this subject has been studied by many authors \cite{lee}. The $s$-wave 
attractions between kaons and nucleons remarkably reduce the
excitation energy of $K^-$ mesons in neutron matter and thereby kaons
should condense when the degeneracy energy of electrons exceeds
it. It is well-known that the kaon condensation results in the large 
softening of equation of state (EOS) and the inducement of new 
processes due to the presence of the condensate.
Since it may occur at
relatively low densities, $(3 - 4)\rho_0$ with $\rho_0$ being the
nuclear density $\simeq 0.16$fm$^{-3}$ in neutron star matter, it is
believed to have many implications on neutron star physics; the fast
cooling mechanism, rapid rotation or the low-mass black hole scenario are
some of them \cite{brow,fujii,bro}. 

We have studied this subject from the point of view of chiral rotation 
on the chiral manifold $G/H$ with 
$G=SU(3)_L\times SU(3)_R, H=SU(3)_V$ \cite{tat}. We have defined the
kaon condensed state by acting a unitary operator $\hat U$ on the normal
matter $|0\rangle$ (the meson vacuum); 
$|K\rangle=\hat U_K(\langle\theta\rangle)|0\rangle$
with $\hat U_K(\langle\theta\rangle)=\exp(i\langle\theta\rangle \hat F_4^5)$,
where $\hat F_4^5$ is the 4-th component of the axial-vector charge of the 
$SU(3)_L\times SU(3)_R$ algebra and $\langle\theta\rangle$ is the
chiral angle. 
\footnote{The notation $\langle\theta\rangle$ implies the thermal
average of $\theta$.}
Since the kaon condensed state is a chirally rotated one, the
change of the kaon-nucleon dynamics from that in normal matter arises from
the symmetry breaking terms in the Hamiltonian. 
Thus we have succeeded in extracting the essential features of kaon
condensation in this formalism.

So far almost all the studies have been done within the mean-field
theory, the classical approximation.
Recently there begin to appear studies about kaon condensation in
protoneutron stars, where thermal effects or roles of the neutrino
trapping become important and we must treat it beyond the mean-field
level \cite{pra}. Low-mass black hole, if exists,
may be produced in this era \cite{bro}. A numerical simulation based
on the general relativity has been already done about the delayed
collapse of protoneutron stars to the low-mass black hole \cite{bau}.
However, they used a cold EOS for the kaon condensed phase, while
temperature $(T)$ rises to several tens MeV there.
Also, in this era, various
time-scales become important; collapsing, neutrino-trapping or  
initial cooling time-scale may be relevant. Hence we cannot assume
chemical equilibrium {\it a priori} for such hot matter since it 
may be achieved through {\it weak} interactions. A problem how
condensate appears and grows to establish chemical equilibrium 
is an interesting
subject to pursure \cite{mut}.  Prakash et al. treated the
kaon-condensed phase at finite temperature and discussed the
properties of protoneutron stars within the meson exchange model,
since there is no consistent theory based on chiral symmetry \cite{pra}.

In a recent paper, Thorsson and Ellis have tried to include quantum or thermal
fluctuations within the heavy-baryon chiral
perturbation theory (HBCPT) \cite{jen}, frozing dynamical degrees of
freedom of nucleons to the thermodynamic potential\cite{tho}. 
However, they have discussed only the
effect of the zero-point fluctuation of kaons at $T=0$ to the one-loop 
order. 
The most interesting result in their paper
may be the dispersion relation of kaons in the condensed phase, which 
is a key ingredient for the discussion of
thermal properties of the condensed phase. Moreover, it plays an
important role in the discussion of the relaxation process from the 
nonequilibrium state to the kaon condensed state in equilibrium 
\cite{mut}. They
have used the Kaplan-Nelson Lagrangian, where the basic variable 
is the $SU(3)$ matrix 
specified by the eight Goldstone fields $\phi_a$,  
$U(\phi_a)=\exp(2iT_a\phi_a)$ with $T_a$ being the generators of
$SU(3)$ Lie algebra. They picked up only the degrees of freedom for
the charged kaon fields $K^\pm$, namely $\phi_4$ and $\phi_5$, and
separated the classical field (the condensate) and the fluctuation
field around it in the standard manner, $K^\pm=\langle
K^\pm\rangle+\tilde K^\pm$. The resultant dispersion relation for
kaons has a very complicated form due to the interaction between the
condensate and fluctuations. Accordingly other thermodynamic
quantities also become too complicated to be tractable. Thus there is
no consistent finite temperature calculation for the chiral case.

In ref.\cite{ty} we have presented a new formalism to treat the kaon
condensation at finite temperature within HBCPT.
We, in this paper, extend the formalism to take
into account the dynamical degrees of nucleons as well as kaons \cite{san}.
We introduced
fluctuations around the condensate by acting the chiral transformation 
on the chiral manifold successively; after generating 
fluctuations around the meson vacuum by a chiral transformation
$(\eta)$ we transform 
this state further by another chiral rotation  $(\zeta)$, which
should be the same as the classical one (see Fig.1). 

\begin{figure}[h]
\epsfxsize=0.3\textwidth
\centerline{\epsffile{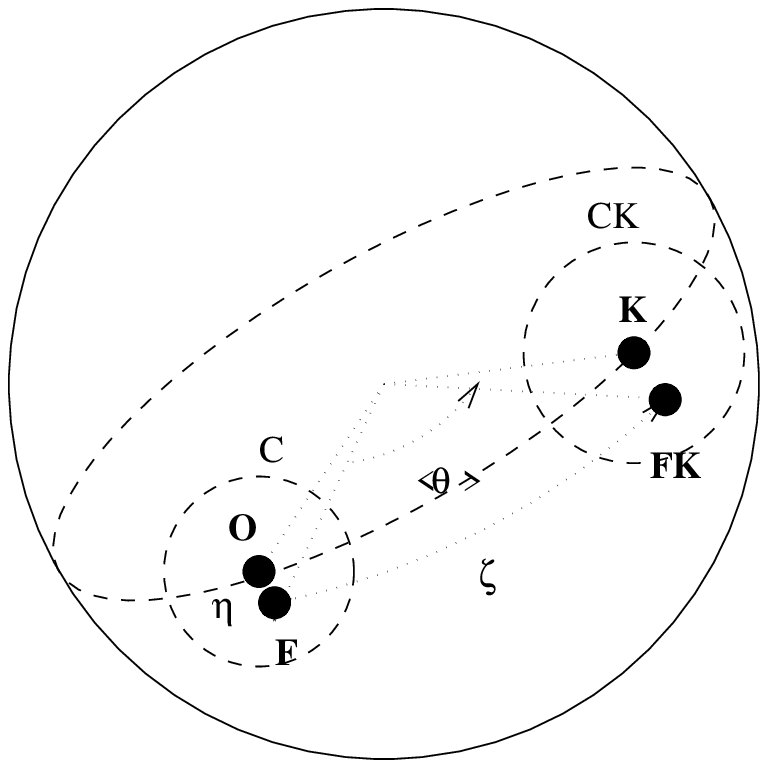}}
\vspace*{5mm}
\caption{Chiral rotations on the $SU(3)$ manifold. Circles $C, CK$
indicate the fluctuation ``areas'' around the vacuum $(O)$ and kaon
condensed state $(K)$, respectively.}
\end{figure}

Since the
constant condensate can be considered as a kind of zero-mode, this
procedure can be regarded as a separation of the zero-mode from
the original matrix $U(\phi_a)$. 
Similar procedure has been applied in the different context 
to see the finite-volume effects of the chiral symmetry 
restoration \cite{gass}.
Then we can obtain the
dispersion relation for the kaon excitation around the condensate.
We have
seen that the dispersion relation shows the similar feature to the one 
by Thorsson
and Ellis, whereas its form is very different. We have also proposed an
approximation which makes the expressions of the thermodynamic
quantities considerably simple. 
We have checked numerically 
that this approximation may work very well. 

In this paper we treat the dynamical
degrees of freedom of kaons and nucleons consistently; since nucleons interact
with kaons, their single-particle energies are modified not only by 
the condensate but also by the fluctuations. For this purpose 
we use the Hartree
approximation for the four-point vertices among kaons and
nucleons; the nucleon Green function contains kaon loops and the kaon
Green function does nucleon loops. 
Therefore we sum up the infinite series of bubble 
diagrams composed 
of one-loop diagrams by kaons and nucleons in a self-consistent way. 
We shall see that our formalism makes the
analysis of the structure of kaon-nucleon ($KN$) dynamics transparent and
clarifies physics included there.

Generally these loop diagrams include divergences, which should exist
even at $T=0$. We discuss how finite contributions can be extracted in 
our formalism.
Analyzing the structure of the self-energy terms, we show these
divergences are properly renormalized to the one-loop order by
introducing relevant counterterms corresponding to the nucleon and kaon
masses, and the $KN$ and $KK$ interaction vertices. Then no more
counterterms are needed to renormalize the thermodynamic
potential. Effects of the zero-point (quantum) fluctuation by kaons
has been shown to be small \cite{tho,ty}. We shall see 
that the zero-point fluctuation by nucleons has a sizable effect on
the ground-state property of the condensed phase.

In \ref{path} we present our formalism to treat fluctuations around the
condensate by way of the path integral approach. In \ref{the} we develop 
our approach and perform the one-loop
calculation. There we get the dispersion relation for kaons which is a 
key object to discuss the properties of the kaon condensed phase at finite 
temperature. We discuss in \ref{eff} how the thermodynamic potential
is renormalized by analyzing the self-energy terms. We also give some
numerical results about the zero-point fluctuation by nucleons.
\ref{sum} is 
devoted to some discussions and concluding remarks. In Appendix A we
discuss the separation of zero-modes, one of which is the would-be
condensate. Some results within the mean-field approximation are
summarized in Appendix B. In Appendix C we give some properties of the 
kaon Green function and expressions of the loop contributions.

\section{Path integral formulation}
\label{path}
\subsection{Partition function}
We start with the Kaplan-Nelson Lagrangian as a chiral
Lagrangian \cite{geo},
\beq
{\cal L}_{chiral}={\cal L}_0+{\cal L}_{SB},
\label{ba}
\eeq
where ${\cal L}_0$ is the symmetric part,
\beq
{\cal L}_0=\frac{f^2}{4}{\rm tr}(\partial_\mu U^\dagger\partial^\mu U)
+i{\rm tr}\{\bar B (\not D-m_B) B\}
+D{\rm tr}(\bar B\gamma_\mu\gamma_5\{A^\mu, B\})
+F{\rm tr}(\bar B\gamma_\mu\gamma_5[A^\mu, B]).
\label{bb}
\eeq
and ${\cal L}_{SB}$ the symmetry breaking part;

\begin{eqnarray}
{\cal L}_{SB}&=&v{\rm tr}\hat m_q(U+U^\dagger-2) \nonumber\\
&+& a_1{\rm tr}\bar B(\xi \hat m_q\xi+ h.c.)B
+a_2{\rm tr}\bar BB(\xi\hat m_q\xi+h.c.)
+a_3\{{\rm tr}\bar BB\}{{\rm tr}(\hat m_q U+h.c.)}
\label{bc}
\end{eqnarray}
with the quark mass-matrix, $\hat m_q\simeq{\rm diag}(0,0,m_s)$. The
coefficient $v$ implies the vacuum expectation value of the quark
bilinear, $-v=\langle 0|\bar q q|0\rangle$, and the 
coefficients $a_i$ measure the strength of the explicitly
symmetry breaking: the kaon mass is given as $m_K^2\simeq vm_s/f^2$
and the $KN$ sigma terms as  
$\Sigma_{Kp}=-m_s(a_1+a_2+2a_3)$ and $\Sigma_{Kn}=-m_s(a_2+2a_3)$. 
$U$ is the $SU(3)$ matrix parametrized by the eight
Goldstone fields $\phi_a$,
\beq
U =\exp[2iT_a\phi_a/f]\in G/H 
\label{bd}
\eeq
and $V_\mu, A_\mu$ are the vector and axial-vector fields constructed
by $U$,
\begin{eqnarray}
V_\mu&\equiv& 1/2(\xi^\dagger\partial_\mu\xi+\xi\partial_\mu\xi^\dagger)
\nonumber\\
A_\mu&\equiv& i/2(\xi^\dagger\partial_\mu\xi-\xi\partial_\mu\xi^\dagger),
\label{be}
\end{eqnarray}
with $U=\xi^2$.
The baryon octet $B$ can be represented by the $SU(3)$ matrix,
\[
B=\left[
\begin{array}{ccc}
\frac{\Sigma^0}{\sqrt{2}}+\frac{\Lambda^0}{\sqrt{6}} &\Sigma^+ &p\\
\Sigma^- &-\frac{\Sigma^0}{\sqrt{2}}+\frac{\Lambda^0}{\sqrt{6}} &n\\
\Xi^- &\Xi^0  &\frac{-2\Lambda^0}{\sqrt{6}}
\end{array}
\right].
\]
Then the  covariant derivative $D_\mu$ is defined by the vector field
$V_\mu$;
\beq
D_\mu B\equiv \partial_\mu B+[V_\mu, B].
\eeq
The transformation properties of the fields $U$ and $B$
under  $G$ are found in, e.g., ref.\cite{geo}.

To treat thermal and quantum fluctuations consistently we use here the path-integral 
formulation. It is well-known that chemical potentials $\mu_a$, which
should be taken into account to
ensure various conservation laws in the ground state,  can be
introduced by ``gauging'' the Lagrangian with artificial ``gauge fields''
$A_\mu\equiv T_aA_a^\mu$ with $A_a^\mu=(\mu_a,{\bf 0})$. 
Here we must take into account conservation of two quantities:
electromagnetic charge and baryon number. Accordingly we consider two
chemical potentials, the charge chemical potential and baryon chemical 
potential.
For the charge chemical potential $\mu_K$,
which we shall see to be identified as the kaon chemical potential,
the ``covariant derivative'' ${\cal D}_\mu U$ for the Goldstone field 
then reads 
\beq
{\cal D}_{\mu=0} U=\partial U/\partial t+i\mu_K[T_{em}, U],\quad 
{\cal D}_\mu U=\partial_\mu U \quad {\rm for \quad others}, 
\eeq
with the charge operator, $T_{em}\equiv T_3+1/\sqrt{3}T_8
={\rm diag}(2/3, -1/3,-1/3)$. For 
baryon field $B$, we must introduce another Abelian gauge field to
ensure the conservation of baryon number, $B_\mu=(\mu_n, {\bf 0})$
with the baryon chemical potential $\mu_n$, which we shall see to be the
neutron chemical potential. Then the ``covariant derivative'' $\nabla_{\mu=0}
B$ for the baryon field reads
\beq
\nabla_{\mu=0} B=\partial B/\partial t-i\mu_n B+i\mu_K [T_{em}, B]
+[V_0,B],\quad 
\nabla_\mu B=D_\mu B \quad {\rm for \quad others}.
\eeq
Therefore the partition function in the imaginary-time formalism can
be represented as follows ($\tau=it, \beta=1/T$);
\beq
  Z_{chiral}=N\int [dU][dB][d\bar B] \exp[S_{chiral}']
\label{part}
\eeq
\beq
S_{chiral}'=\int_0^\beta d\tau\int d^3x
{\cal L}_{chiral}\left({\cal D}_\mu U, U, \nabla_\mu B, B\right)
=\int_0^\beta d\tau\int d^3x[{\cal L}_{chiral}+\delta
{\cal L}], 
\eeq
where the ``time-derivative'' should be read as the derivative with
respect to imaginary time.
Here we can see that there appears an additional term $\delta {\cal L}$ 
besides the original chiral Lagrangian ${\cal L}_{chiral}$, which is
in general {\it non-invariant} under the chiral transformation. 
For the Lagrangian (\ref{ba}), we find 
\begin{eqnarray}
\delta {\cal L}&=& -\frac{f^2\mu_K}{4}{\rm tr}\{[T_{em}, U]
\frac{\partial U^\dagger}{\partial\tau}+
\frac{\partial U}{\partial\tau}[T_{em}, U^\dagger]\}
-\frac{\mu_K}{2}{\rm tr}
\{B^\dagger[(\xi^\dagger[T_{em}, \xi]+\xi[T_{em}, \xi^\dagger]), B]\}
\nonumber\\
&-&\frac{f^2\mu_K^2}{4}{\rm tr}\{[T_{em}, U][T_{em}, U^\dagger]\}
+\mu_n{\rm tr}\{B^\dagger B\}-\mu_K{\rm tr}\{B^\dagger[T_{em},B]\} .
\end{eqnarray}
Here the first term corresponds to the mesonic charge, 
and the last two terms indicate nothing else but
the baryon number and the electromagnetic charge of baryons.

\subsection{Transformation of the coordinates}
We show our formulation to include thermal and quantum fluctuations in
the chiral Lagrangian. Before that let us recall the tree case.
We have described the kaon condensed state $|K\rangle$ , which 
is specified by the chiral angle $\langle\theta\rangle$, the order
parameter of kaon condensation,
as a chiral-rotated state \cite{tat};
\beq
|K\rangle=\hat U_K(\langle\theta\rangle)|0\rangle
\label{state}
\eeq
with the operator 
$\hat U_K(\langle\theta\rangle)=\exp(i\hat
F_4^5\langle\theta\rangle)$, where $\hat F_4^5$ is
the 4-th element of the axial-vector charge in $SU(3)_L\times SU(3)_R$ 
algebra and $|0\rangle$ represents the normal nuclear
matter. We can also describe this point on the 
$SU(3)_L\times SU(3)_R/SU(3)_V\simeq SU(3)$ manifold which is
coordinated by the Goldstone fields, 
$U(\phi_a)=\exp(2iT_a\phi_a/f)\in SU(3)$ and the meson vacuum $|0\rangle$
corresponds to the fiducial point on the manifold by the spontaneous
symmetry breaking.
Then kaon condensed state Eq. (\ref{state}) corresponds to the 
$SU(3)$ matrix $U_K=\exp[2i\{T_4\langle\phi_4\rangle
+T_5\langle\phi_5\rangle\}/f]$, and can be represented as a
chiral rotated one from the vacuum on this manifold,
\beq
U_K(\langle\theta\rangle)=\zeta U_V\zeta=\zeta^2
\label{zuvz}
\eeq 
where  $U_V=U(\phi_a=0)=1$ corresponds to the meson vacuum and $\zeta$ 
is the operator, $\zeta=\exp(i\langle M\rangle/\sqrt{2}f)$ with the 
constant matrix
\[
M=\left[
\begin{array}{ccc}
0 & 0 & K^+ \\
0 & 0 & 0\\
K^- & 0 & 0
\end{array}
\right]
,\qquad K^{\pm}=(\phi_4\pm i\phi_5)/\sqrt{2}\qquad {\rm and}\qquad
\theta^2\equiv 2K^+K^-/f^2. \hskip 3cm
\]
Thus we have seen that the classical kaon field $\langle K^\pm\rangle$ 
induces the chiral rotation on the manifold \cite{tat}. 

\begin{figure}[h]
\epsfxsize=0.6\textwidth
\centerline{\epsffile{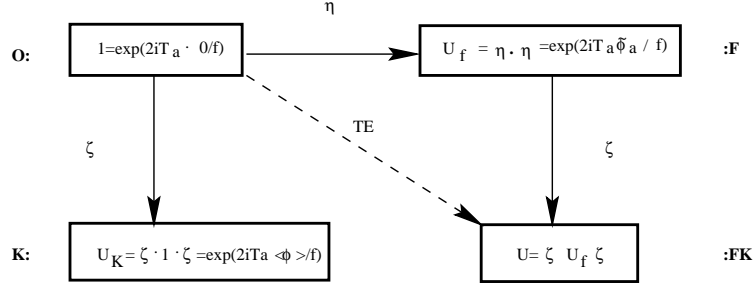}}
\vspace*{5mm}
\caption{Diagram showing the relation among the transformations 
on the chiral manifold (c.f. Fig.~1). The symbol $TE$ indicates 
the one by Thorsson and Ellis [13].}
\end{figure}

When we introduce the quantum or thermal fluctuation, there
may be several ways (Figs.1,2). 
Thorsson and Ellis introduced the kaon fluctuation
field $\tilde K^\pm $ as a deviation from the classical kaon field in
the standard manner,
\footnote{Hereafter, we omit the tilde on the field for the sake of 
simplicity.}
\beq
K^\pm=\langle K^\pm\rangle+\tilde K^\pm. 
\eeq
Here we introduce the fluctuation fields by developping the idea of
chiral rotation mentioned above. First, we notice that any
fluctuation 
of the Goldstone
fields can be introduced by a chiral rotation $U_f$ from the vacuum $U_V=1$,
\beq
\eta: \qquad U_f=\eta U_V \eta=\eta^2
\label{vacchiral}
\eeq
with $\eta=\exp(iT_a\phi_a/f)$. The subsequent chiral rotation by
$\zeta$ transforms $U_f$ into the form,
\beq
\zeta: \qquad U(\langle\theta\rangle, \phi_a)=\zeta U_f(\phi_a)\zeta.
\label{ansatz}
\eeq 
We can easily see that in the limit $\phi_a\rightarrow 0$, which
corresponds to the previous classical approximation, the matrix $U$
is reduced to $U_K$ (\ref{zuvz}).
Thus we have introduced the fluctuation fields by the two-step
procedure. 
\footnote{The similar decomposition was also used in the context of
the loop expansion within chiral perturbation theory \cite{san} or the 
bound-state approach to the Skyrme model \cite{cal}.}
It is to be noted that the constant component in $\phi_a$ is redundant 
since it can be always absorbed into $\zeta$ by redefinition.
It may be also worth noting that we can easily take into account 
fluctuations of {\it any} meson
field in the kaon condensed state by the form of Eq.~(\ref{ansatz}).
\footnote{Alternatively, we can regard this procedure as a separation 
of the ``zero-mode'' from the full $SU(3)$ matrix $U(\phi_a)$ (see
Appendix A). 
Using Eq.~(\ref{extrem}) we can see that the value of the
thermal average $\langle\theta\rangle$ should be determined by
imposing the extremum condition for the resultant thermodynamic
potential at the end.}
Accordingly,
$\xi$ operator, which is defined by $U=\xi^2$, can be obtained
by solving the subsidiary equation,
\beq
\xi=\zeta \eta u^\dagger=u \eta \zeta,
\label{sub}
\eeq 
where  the matrix $u$ is defined by the second equality in Eq.(\ref{sub}). 
It depends on $\phi_a, 
\langle\theta\rangle$ in a complicated nonlinear way, and thereby it is 
very difficult to
find $u$ for the general form of $U_f$ \cite{geo}. A systematic way to 
find the form of $u$ may be the perturbative method: expanding the
matrices $\eta, u$ with respect to the fluctuation field $\phi_a$, we
can solve Eq.~(\ref{sub}) order by order. Hence we can find the
relevant form
of $u$ up to any order of $\phi_a$, depending on the problem or the
approximation we are interested in.
On the other hand, if we 
restrict ourselves to the fluctuations of kaon sector s.t.
\beq
U_f=\exp(i\sqrt{2}M/f),
\eeq
then we can find $u$ in the closed form,
\beq
u={\rm diag}(\kappa^*/|\kappa|, 1, \kappa/|\kappa|),
\label{solu}
\eeq
with  
\beq
\kappa=\cos(\langle\theta\rangle/2)\cos(\theta/2)
-\frac{K^+\langle K^-\rangle}{|\langle K\rangle||K|}
\sin(\langle\theta\rangle/2)\sin(\theta/2).
\label{soluk}
\eeq
Here the following relation is to be noted,
\beq
u^\dagger T_{em}u=T_{em}.
\label{utu}
\eeq

Defining a new baryon field $B'$ by the use of  the matrix $u$,
\beq
 B'=u^\dagger B u,
\label{bp}
\eeq
we can see that the chiral-invariant pieces of the Lagrangian are
not changed and only the symmetry breaking pieces play essential roles,   
\begin{eqnarray}
{\cal L}_{chiral}(U_K, B)
={\cal L}_0(U_K, B)+{\cal L}_{SB}(U_K, B)&\longrightarrow&
{\cal L}_0(U_f, B')+{\cal L}_{SB}(\zeta U_f\zeta,u B'
u^\dagger)\nonumber\\
\delta{\cal L}(U_K, B)&\longrightarrow&\delta{\cal L}(\zeta U_f\zeta,u B'
u^\dagger).
\end{eqnarray}
Thus all the dynamics of kaons and baryons in the condensed
phase are completely prescribed by the non-invariant terms under
chiral transformation.
It is to be noted that this feature is quite similar to the one within the
classical approximation at $T=0$ \cite{tat}.

Using Eqs.~(\ref{solu}), (\ref{bp}), we find
\begin{eqnarray}
\delta{\cal L}(\zeta U_f\zeta,u B'
u^\dagger)&=&-(\cos(\langle\theta\rangle-1)\mu_K {\rm tr}\{B'^\dagger[V_3,B']\}
+f^2\mu_K^2/2\sin^2\langle\theta\rangle\nonumber\\
&-&\mu_K\cos\langle\theta\rangle(K^+\dot K^--\dot K^+
K^-)+\mu_K\cos\langle\theta\rangle K^+ K^-/f^2{\rm tr}\{B'^\dagger[V_3,B']\}
\nonumber\\
&+&\cos^2\langle\theta\rangle\mu_K^2K^+
K^--\mu_K^2/4\sin^2\langle\theta\rangle(\exp(i\gamma)K^-+\exp(-i\gamma)K^+)^2
\nonumber\\
&+&\mu_n{\rm tr}\{B'^\dagger B'\}-\mu_K{\rm tr}\{B'^\dagger[T_{em},B']\}...,
\end{eqnarray}
with the V-spin operator, $V_3=1/2(T_3+\sqrt{3}T_8)$, 
and $\dot K^\pm=\partial K^\pm/\partial \tau$, where the phase 
$\gamma$ is defined by 
$\exp(\pm i\gamma)=\langle K^\pm\rangle/|\langle K\rangle|$, 
which can be absorbed
into the fluctuation field by redefining it, $\exp(\mp i\gamma)K^\pm
\rightarrow K^\pm$. In other words the phase of the condensate is
arbitrary, reflecting $V_3$ symmetry in the chiral Lagrangian, and 
we can choose $\gamma=0$ 
without loss of generality. As a consequence $V_3$ symmetry is 
spontaneously broken in the condensed phase, and
thereby we shall see the Goldstone mode associated with this symmetry
(see \ref{eff}).
\footnote{We met the similar situation in the context of pion
condensation \cite{cam}.} 

If ${\cal L}_{SB}$ originates from the quark mass term 
, ${\cal L}_{mass}=-\bar q\hat m_q q$, we can further explore the
transformation property of ${\cal L}_{SB}$ in a model-independent way: 
first we can see 
\beq
{\cal L}_{mass} \rightarrow {\cal L}_{mass}
-i\sin\theta[\hat F_4^5, {\cal L}_{mass}]+(\cos\theta-1)\Sigma_K
\label{masstrans}
\eeq
with the sigma term $\Sigma_K=[\hat F_4^5,[\hat F_4^5,{\cal
L}_{mass}]]$, 
under the transformation by $\eta$ in Eq.~(\ref{vacchiral}). 
Similarly, ${\cal L}_{mass}$
transforms like (\ref{masstrans}) with $\langle\theta\rangle$ instead of 
$\theta$ under the chiral rotation $\zeta$ in Eq.~(\ref{ansatz}). 
Hence ${\cal L}_{mass}$ eventually transforms under the 
successive chiral 
transformations (\ref{vacchiral}) and (\ref{ansatz}) like
\beq
{\cal L}_{mass} \rightarrow {\cal L}_{mass}
-i\left(\sin(\langle\theta\rangle+\theta)-\sin\theta\right)
[\hat F_4^5, {\cal L}_{mass}]
+\left((\cos(\langle\theta\rangle+\theta)-\cos\theta\right)\Sigma_K,
\eeq
where we have used the property $[\hat F_4^5,[\hat F_4^5,[\hat F_4^5,{\cal
L}_{mass}]]]=[\hat F_4^5, {\cal L}_{mass}]$. The second term is a 
parity-odd operator and irrelevant due to the fact, 
$\langle \bar q\gamma_5 q\rangle=0$. The term proportional to
$\sin\langle\theta\rangle$ included in the last term is also 
irrelevant in the following
calculation bacause of the thermal-equilibrium condition 
(c.f. Eq.~(\ref{extrem})). Therefore there is essentially left the term,
\beq
{\cal L}_{mass} \rightarrow {\cal L}_{mass}
+(\cos\langle\theta\rangle-1)\cos\theta\Sigma_K.
\eeq
Accordingly, in terms of the effective Lagrangian ${\cal L}_{SB}$, it
should transform like 
\beq
{\cal L}_{SB}(\zeta U_f\zeta,u B'u^\dagger)
={\cal L}_{SB}(U_f, B')
+(\cos\langle\theta\rangle-1)\cos\theta\Sigma_K(1, B')
\label{sbtrans}
\eeq
besides the irrelevant terms, where $\Sigma_K(1, B')$ has a standard form,
\beq
\Sigma_K(1, B')=f^2m_K^2-\Sigma_{Kp}\bar p'p'-\Sigma_{Kn}\bar n' n',
\label{symbr}
\eeq
for nuclear matter.
\footnote{We, hereafter, omit the prime on the nucleon fields for simplicity.}
In Eq.~(\ref{symbr}) we have used the following notations:
$f^2m_K^2=v[m_u+m_s], \Sigma_{Kp(n)}=\langle
p(n)|\Sigma_K|p(n)\rangle$. This consequence is rather general and,   
of course, consistent with that in the
Kaplan-Nelson Lagrangian (\ref{bc}).

For the integration measure $(N=3)$, it is given as
\beq
[dU]=M(\phi)\prod_{i=1}^{N^2-1}[d\phi_i],
\label{measure}
\eeq
where $M(\phi)$  is the Lee-Yang term \cite{hon}, $M(\phi)
=\exp\left[ \frac{1}{2}\delta^{(4)}(0)\int_0^\beta d\tau\int d^3x\ln 
g(\phi)\right]$ with the Cartan-Killing metric on the $SU(N)$ manifold
from the invariant line element,
$
ds^2=g_{ik}(\phi)d\phi_id\phi_k=1/2{\rm tr}[dUdU^\dagger]
$. It is to be noted that 
the Lee-Yang term is important to make the integration measure chiral 
invariant.
\footnote{In ref.\cite{tho}, their treatment of the Lee-Yang term is
obscure: $\gamma^{1/2}$ in Eq~(13) in ref.\cite{tho}, 
which corresponds to $g$ in
our notation, should read 
$\exp\left[ \frac{1}{2}\delta^{(4)}(0)\int_0^\beta d\tau\int d^3x\ln 
\gamma\right]$.}
The measure (\ref{measure}) is invariant under the
transformation (\ref{ansatz}) by $\zeta$,
\beq
[dU]=[d(\zeta U_f\zeta)]=[dU_f],
\eeq
because $G$ is the isometry, ${\rm tr}[\zeta dU_f\zeta \zeta^\dagger
dU_f^\dagger\zeta^\dagger]
={\rm tr}[dU_fdU_f^\dagger]$, so that the geometry around the
condensed point is the same as in the vicinity of the vacuum.
Since we are concerned with the fluctuations, the measure can be
approximated by the one with flat curvature \cite{gass}, 
$[dU_f]\simeq \prod_{i=1}^{N^2-1}[d\phi_i]$. 
In this case we are not worried by the Lee-Yang term because of $g= 
1+O(\phi_i^2/f^2)$. Thus we find
\beq
Z_{chiral}\simeq \int\prod_{i=1}^{N^2-1}[d\phi_i]
[dB'][d\bar B']\exp[S_{chiral}^{eff}(\zeta, U_f, B')].
\eeq

\section{Thermodynamic potential}
\label{the}
\subsection{Partition function for a chiral Lagrangian}
Using the formulation given in \ref{path} we evaluate the partition function 
in the kaon condensed state.
\footnote{We briefly discuss the case of the classical
approximation in Appendix B, where we can see that our previous
results given in ref.\cite{tat} are recovered in a model-independent way.}
In the following calculation we only consider the one-loop diagrams,
and thereby we retain only the quadratic terms with respect to the
kaon field. Moreover, we are concerned here with only charged-kaon
fluctuations and nuclear matter without
hyperons.
\footnote{An extension to include hyperons may be straightforward, 
but note that there is
still large ambiguity about their coupling strength with hadrons,
e.g. see ref.~\cite{pra}.}   
Then we can write
\beq
Z_{chiral}=N\int [d\phi_4][d\phi_5][d\psi^\dagger][d\psi]
\exp[S^{eff}_{chiral}]
\label{zchiral}
\eeq
with the spinor 
$\psi=\left(
\begin{array}{c}
p\\
n
\end{array}
\right)
$ by the use of Kaplan-Nelson Lagrangian (\ref{ba}).
\beq
S^{eff}_{chiral}=S_c+S_K+S_N+S_{int},
\eeq
where 
\beq
S_c=\beta
V\left[\frac{1}{2}\mu_K^2f^2\sin^2\langle\theta\rangle
-f^2m_K^2(1-\cos\langle\theta\rangle),
\right]
\eeq
\begin{eqnarray}
S_K=\int_0^\beta d\tau\int
d^3x&& \left [\partial_\mu K^+\partial_\mu K^-
-\cos\langle\theta\rangle (m_K^2-\mu_K^2\cos\langle\theta\rangle)K^+ K^- 
-\mu_K\cos\langle\theta\rangle K^+
\stackrel{\leftrightarrow}{\partial_\tau} K^-\right.\nonumber\\
& &
-\frac{\mu_K^2}{4}\sin^2\langle\theta\rangle(K^++K^-)^2
+O(|K|^4)\Bigl.\Bigr ],
\label{sk}
\end{eqnarray}
with the shorthand notation 
$\partial_\mu K^+\partial_\mu K^-=-\partial_\tau K^+\partial_\tau K^--
\mbox{\boldmath$\nabla$} K^+ \mbox{\boldmath$\nabla$} K^- $,
\begin{eqnarray}
S_N=\int_0^\beta d\tau\int
d^3x&&\left[\bar\psi(-\gamma^0\frac{\partial}{\partial\tau}
+i\mbox{\boldmath$\gamma$}\cdot\mbox{\boldmath$\nabla$}-M
+\mu_N\gamma^0\right .\nonumber\\
&+&\left .\mu_K(1-\cos\langle\theta\rangle)\gamma^0\frac{3+\tau_3}{4}+
(1-\cos\langle\theta\rangle)\Sigma_{KN})\psi\right],
\label{sn}
\end{eqnarray}
and $S_{int}$ is the interaction term, which mainly consists of the
two terms, the sigma-term and the Tomozawa-Weinberg (TW) term,
\begin{eqnarray}
S_{int}=\int_0^\beta d\tau\int d^3x\Bigl [\Bigr.
&&\frac{1}{f^2}\cos\langle\theta\rangle\bar\psi\Sigma_{KN}\psi K^-
K^+\nonumber\\
&-&\frac{1}{f^2}\bar\psi\gamma_0\frac{3+\tau_3}{8}\psi
\left\{K^+\stackrel{\leftrightarrow}{\partial_\tau} K^-
-2\mu_K\cos\langle\theta\rangle K^+K^-\right\}+O(|K|^4)
\Bigl.\Bigr]
\label{sint}
\end{eqnarray}
with the $KN$ sigma term $\Sigma_{KN}$. It is to be noted that the 
sigma-term $\Sigma_{KN}$ and
nucleon chemical potential $\mu_N$ are $2\times 2$ diagonal matrices in the
isospin space: 
$\Sigma_{KN}=\left(\begin{array}{cc} \Sigma_{Kp} & 0 \\0 & \Sigma_{Kn}
\end{array}\right)$ and
$\mu_N=\left(\begin{array}{cc} \mu_p & 0 \\0 & \mu_n
\end{array}\right)$ with $\mu_p\equiv \mu_n-\mu_K$, respectively.


Since the effective action $S_{chiral}^{eff}$ is already bilinear with
respect to the nucleon field, we can easily integrate out the nucleon
degree of freedom in Eq.~(\ref{zchiral}),
\beq
Z_{chiral}=N\int [d\phi_4][d\phi_5]
\exp[S'_K]{\rm Det} G=N\int [d\phi_4][d\phi_5]\exp[
S'_K+{\rm Tr}\ln G].
\label{zch}
\eeq
with $ \tilde S'_K=S_c+S_K$ and the Dirac operator $G$. 
The capital letter for determinant or trace is
used for the Dirac operator $G$ of infinite dimension over 
space-imaginary-time. 
The Dirac operator $G$ is composed of two parts, $G=G_0+\delta G$,
with the ``free'' part 
\beq
G_0=-\frac{\partial}{\partial\tau}
+i\gamma_0 \mbox{\boldmath$\gamma$}\cdot\mbox{\boldmath$\nabla$}-M\gamma_0
+\mu_N
+\mu_K(1-\cos\langle\theta\rangle)\frac{3+\tau_3}{4}+
(1-\cos\langle\theta\rangle)\gamma_0\Sigma_{KN},
\eeq
and the interaction part
\beq
\delta G=\frac{\Sigma_{KN}}{f^2}\cos\langle\theta\rangle\gamma_0 K^-
K^+
-\frac{1}{f^2}\frac{3+\tau_3}{8}
\left\{K^+\stackrel{\leftrightarrow}{\partial_\tau} K^-
-2\mu_K\cos\langle\theta\rangle K^+K^-\right\}+O(|K|^4).
\eeq

A naive 
treatment may proceed with treating $G_0$ as the zeroth-order operator and 
expanding ${\rm Tr}\ln G$ in terms of $\delta G$,
\beq
{\rm Tr}\ln G={\rm Tr}\ln G_0+{\rm Tr}[G_0^{-1}\delta G]+O(|K|^4).
\label{trlng}
\eeq
The first term gives a one-loop contribution of nucleons to the
partition function, and the second term to 
the kaon self-energy term. The latter, together with $S_K$, gives a
one-loop contributions of kaons to the partition function.
However, this procedure is not adequate for treating the $KN$
interactions in a consistent way; a simple
analysis of the loop counting shows that the kaon one-loop
contribution includes an infinite series of nucleon loops, while the
nucleon one-loop contribution does not include any kaon loop. 
In the following
we use the Hartree approximation to take into account all the bubble
diagrams by both nucleons and kaons non-perturbatively, while
the exchange diagrams are discarded. Generally speaking, the exchange
diagrams are suppressed in the high-density or high-temperature limit,
compared with direct diagrams (see Fig.3  for the three loop case)
\cite{kap,dol}. 

\begin{figure}[h]
\epsfxsize=0.5\textwidth
\centerline{\epsffile{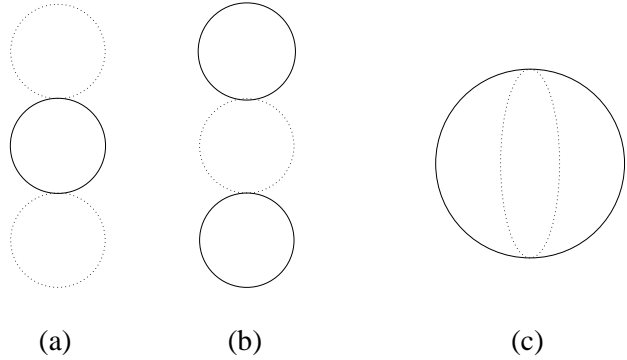}}
\vspace*{5mm}
\caption{Three-loop diagrams. (a) and (b) are the direct(bubble)
diagrams, which are taken into account by
the Hartree approximation. (c) is the exchange diagram, which should
be suppressed by the power of density or temperature in comparison
with (a) or (b).}
\end{figure}

We define a new operator $G_H$, $
G_H=G_0+\langle\delta G\rangle$ by taking the thermal avarage of $\delta 
G$ for kaons, 
\beq
G_H=-\frac{\partial}{\partial\tau}
+i\gamma_0 \mbox{\boldmath$\gamma$}\cdot\mbox{\boldmath$\nabla$}
+\mu_N+\frac{3+\tau_3}{8}
(n_K^2+2\mu_K(1-\cos\langle\theta\rangle))
-M^*\gamma^0
\label{gh}
\eeq
with the effective mass of
the nucleon, $M^*=\left(\begin{array}{cc} M^*_p & 0 \\0 & M^*_n
\end{array}\right)$ with
\beq
M^*_i=M-\Sigma_{Ki}/f^2\cos\langle\theta\rangle n_K^1
-\Sigma_{Ki}(1-\cos\langle\theta\rangle). 
\label{effm}
\eeq
The quantities 
$n_K^1, n_K^0$ mean the thermal kaon-loop contributions for the
self-energy of nucleons and they can be represented in terms of the
thermal Green function of kaons (see Appendix C)
\footnote{The quantity $n_K^1$ includes the divergence and should
be properly renormalized. We shall meet other quantities that also
include some divergences in this section.  
We leave the renormalization problem in \ref{eff} 
and hereafter treat them schematically in this section.}
; 
$n_K^1\equiv \langle K^+K^-\rangle$ and 
$n_K^2\equiv -n_K^0
+2\mu_K\cos\langle\theta\rangle n_K^1$ with 
$n_K^0\equiv \langle K^+\stackrel{\leftrightarrow}{\partial_\tau} K^-\rangle$.

Using $G_H$ we can rewrite Eq.~(\ref{trlng}) as 
\beq
{\rm Tr}\ln G={\rm Tr}\ln G_H+{\rm Tr}[G_H^{-1}\delta G]
              -{\rm Tr}[G_H^{-1}\langle\delta G\rangle]+O(|K|^4).
\label{trlnh}
\eeq
Substituting Eq.~(\ref{trlnh}) into Eq.~(\ref{zch}), we find
\beq
Z_{chiral}=N\int [d\phi_4][d\phi_5]
\exp[\tilde S_K]{\rm Det} G_H,
\eeq
with $\tilde S_K=S_c+S_K^{eff}-{\rm Tr}[G_H^{-1}\langle\delta G\rangle]
+O(|K|^4)$. 
Thus we construct an effective action for the kaon fluctuation field, 
\begin{eqnarray}
S_K^{eff}=\int_0^\beta d\tau\int d^3x&&\left[-\partial_\mu K^+\partial_\mu K^-
+\cos\langle\theta\rangle(\mu_K^2\cos\langle\theta\rangle-m_K^2+\sigma+2b\mu_K)
K^+K^-\right.\nonumber\\
&-&(\mu_K\cos\langle\theta\rangle+b)K^+
\stackrel{\leftrightarrow}{\partial_\tau} K^--\frac{\mu_K^2}{4}\sin^2\langle\theta\rangle(K^++K^-)^2 
+O(|K|^4)
\left .\right],
\label{effl}
\end{eqnarray}
where $\sigma\equiv f^{-2}{\rm Tr}[G_H^{-1}\gamma_0\Sigma_{KN}]$ and $
b\equiv f^{-2}{\rm
Tr}[G_H^{-1}(3+\tau_3)/8]=(4f^2)^{-1}(\rho_n+2\rho_p)$
with nucleon number densities $\rho_i (i=n, p)$. 
It may be worth noting that the self-energy term $\sigma$ 
contains some divergences
(see \ref{eff}), and is reduced to
$\sigma_{tree}\simeq f^{-2}(\Sigma_{Kp}\rho_p+\Sigma_{Kn}\rho_n)$ 
in the static limit of nucleons \cite{ty}.

Expanding the fluctuation field according to
\beq
\phi_{4,5}=\sqrt{\frac{\beta}{V}}\sum_{n,{\bf p}\neq 0}
e^{i({\bf p}\cdot {\bf r}+\omega_n\tau)}\phi_{4,5}(n,{\bf p})
\label{period}
\eeq
with the Matsubara frequency $\omega_n=2\pi n T$, we find
\beq
Z_K^{eff}=\int\prod_{n,{\bf p}\neq 0}[d\phi_4(n,{\bf p})][d\phi_5(n,{\bf p})]
e^{S_K^{eff}}.
\eeq
It is to be noted that the zero mode ($n=0,{\bf p}=0$) should be
removed to avoid the double counting (see Eq.~(\ref{ansatz})).  
Here the effective action reads
\beq
S_K^{eff}=-\frac{1}{2}\sum_{n, {\bf p}\neq 0}(\phi_4(-n,-{\bf p}), 
\phi_5(-n,-{\bf p}))
D^{eff}\left(
\begin{array}{c}
\phi_4(n,{\bf p})\\
\phi_5(n,{\bf p})
\end{array}
\right)
\eeq
with the inverse thermal Green function,
\beq
D^{eff}=\beta^2\left(
\begin{array}{cc}
\omega_n^2+(\tilde\omega_--\tilde\mu_K)
(\tilde\omega_++\tilde\mu_K)+\mu_K^2\sin^2\langle\theta\rangle & 
2(\tilde\mu_K+b)\omega_n\\
-2(\tilde\mu_K+b)\omega_n & 
\omega_n^2+(\tilde\omega_--\tilde\mu_K )
(\tilde\omega_++\tilde\mu_K)
\end{array}
\right),
\eeq
where 
\beq
\tilde\omega_\pm=\pm b+(p^2+C^2)^{1/2}
\eeq
with $C^2=\tilde m_K^{*2}+b^2$ 
and $\tilde\mu_K=\mu_K\cos\langle\theta\rangle$ with the effective mass, 
$\tilde m_K^{*2}=\cos\langle\theta\rangle[m_K^2-\sigma]$.
It is to be noted that the effects of the condensate come in through
the modifications of the chemical potential, 
$\mu_K\rightarrow \mu_K\cos\langle\theta\rangle$, and the effective
mass,$m_K^{*2}=[m_K^2-\sigma]\rightarrow \cos\langle\theta\rangle
m_K^{*2}$, except
the additional term, $\mu_K^2\sin^2\langle\theta\rangle$.  
Then, the partition function can be simply represented as
\beq
\ln Z_{chiral}=S_c-{\rm Tr}[G_H^{-1}\langle\delta G\rangle]-1/2
{\rm Tr}'\ln D^{eff}+{\rm Tr}\ln G_H,
\label{zk}
\eeq
where the symbol ${\rm Tr}'$ means that the zero-mode should be removed.

\subsection{Dispersion relations for kaonic modes}
The excitation energy of the kaonic modes are given by the
solutions $\omega\equiv i\omega_n$, which satisfy $|D^{eff}|\equiv 
{\rm det}(D^{eff})=0$ (see Appendix B):
\beq
\omega^4-2(c_1+c_3+2c_2^2)\omega^2+2c_1c_3+c_1^2=0,
\eeq
where $c_1=(\tilde\omega_--\tilde\mu_K)
(\tilde\omega_++\tilde\mu_K)$, 
$c_2=\tilde\mu_K+b$  
and $c_3=1/2\mu_K^2\sin^2\langle\theta\rangle$.
Then we have two solutions which correspond to the $K^\pm$ modes,
\beq
E_\pm^2=(c_1+c_3+2c_2^2)\pm\sqrt{(c_3+2c_2^2)^2+4c_1c_2^2}.
\label{disp}
\eeq
In the limit $\langle\theta\rangle=0$, $c_1\rightarrow (\omega_--\mu_K)
(\omega_++\mu_K) $, $c_2\rightarrow \mu_K+b$ and $c_3\rightarrow
0$, with the dispersion relation in the normal phase, 
$\omega_\pm({\bf p})=\pm b+(p^2+m_K^{*2}+b^2)^{1/2}$. 
Then $E^2_\pm\rightarrow (\omega_\pm\pm\mu_K)^2$,
which recover the previous results as they should do.

In the condensed phase ($\langle\theta\rangle\ne 0$) 
the mode corresponding to $E_-$ is the Goldstone mode as a consequence 
of the breakdown of V-spin symmerty;
we can choose such  form for the condensate as
$\langle K^\pm\rangle=f\langle\theta\rangle/\sqrt{2}\cdot\exp(\pm i\gamma)$ 
with an arbitrary $\gamma$. However, once a phase is
chosen, 
the kaon-condensed state is a symmetry-broken phase with respect to the
rotation around the third axis in $V$-spin space by the $U(1)$ operator, 
$U_V(\nu)=\exp(iV_3\nu)$ with arbitrary $\nu$, 
while the effective
Lagrangian is still inavariant under the transformation, 
$U \rightarrow U_V U U_V^\dagger$. 
Thus we observe a spontaneous symmetry breaking (SSB) there.

It is easy to show $E_-\sim 0$ for ${\bf p}=0$. In the classical (tree-level)
approximation $c_1({\bf p}=0)=0$ (see Appendix B), 
which directly means $E_-({\bf p}=0)=0$;
actually we can expand $E_-$,
\beq
E_-^2\simeq \frac{c_3}{2C^2}p^2+\frac{p^4}{4C^2}+...,
\label{exp}
\eeq
for small momentum, $|{\bf p}|\ll C$. This dispersion relation may
remind us of the Bogoliubov spectrum for the phonon mode in
superfluidity or Bose-Einstein condensation \cite{lan}.  
When we take into account
the fluctuation effects (kaon loops) in the thermodynamic potential, 
their order of magnitude is $O(\hbar)$.
Accordingly, 
this relation should be modified in 
$O(\hbar)$, because any {\it kaon loop} is not included in
Eq.~(\ref{disp}). 
\footnote{On the other hand, the nucleon loops can be treated
consistently (see \ref{eff}).}
Hence we should find $c_1({\bf p}=0)=0+O(\hbar)$, which
means that once fluctuations are included in the thermodynamic
potential, the soft mode loses the
Goldstone-boson nature. This situation is inevitable in the
perturbative treatment~\cite{kap}: our thermodynamic potential takes into
account the one-loop diagrams, whereas the self-energy diagram for
kaons never include any kaon loop.
However, we may expect its deviation 
to be small. Actually, we
have found that the zero-point contribution from kaon loops 
is very small compared
with tree-level contributions~\cite{tho,ty}.  
When we consider the thermal kaon loop, $E_-$ directly enters into the
Bose-Einstein distribution function, $f_B(E)=[\exp(\beta E)-1]^{-1}$, 
 and it should diverge at ${\bf
p}=0$. The other is the massive mode, $E_+\gg 100$MeV, 
and we may discard it for
temperature we are interested in $(T\leq 100$MeV).
Thus only the thermal $K^-$ loops play an important role due to this
property.

Moreover, if we can safely neglect the $c_3$ term, we get a simple expression
for $E_\pm$,
\beq
E_\pm=\tilde\omega_\pm\pm\tilde\mu_K
=\sqrt{p^2+C^2}\pm(b+\tilde\mu_K).
\label{dispa}
\eeq
There are several reasons to support  the pertinence of these formulae, 
especially for $E_-$. 
First, we can see from Eq.~(\ref{exp}) that in the high momentum limit, 
$|{\bf p}|\gg \sqrt{2c_3}=\mu_K\sin\langle\theta\rangle$, 
the difference of $E_-$ between 
Eq.~(\ref{disp}) and Eq.~(\ref{dispa}) converges to be trifling.
\footnote{This property is very important for the renormalization of kaon
loops, where ultraviolet divergences become relevant, see \ref{eff}.}
On the other hand, in the low momentum
limit, $|{\bf p}|\ll C$, $E_-$ can be expanded as
\beq
E_-\simeq \frac{p^2}{2C}+...
\eeq 
within the tree approximation, which just corresponds to the second term in
Eq.~(\ref{exp}). 
Therefore, 
the approximated formula (\ref{dispa}) may be well justified for
the momentum, $|{\bf p}|\gg \mu_K\sin\langle\theta\rangle$,
while it spoils the Goldstone-boson behavior ($E_-\propto |{\bf p}|$) 
near ${\bf p}=0$. So we must
carefully treat the small-${\bf p}$ region, when we use the approximation.
Comparing (\ref{disp}) and (\ref{dispa}) we can expect that the
``nonrelativistic'' dispersion relation (\ref{dispa}) 
holds under the condition, $v^2\equiv c_3/2C^2\ll 1$, namely the small velocity
limit. Typically $\mu_K\sim O(100{\rm MeV})$ and $C\sim O(m_K)$
\cite{lee}, so that we find $v^2\sim O(10^{-2})\ll 1$.

At finite temperature we can find a more meaningful criterion for the
pertinence of Eq.~(\ref{dispa}).
Consider a typical integral at finite temperature,
\beq
I=\int\frac{d^3 p}{(2\pi)^3}f_B(E_-({\bf p})),
\eeq
which is the particle number of $K^-$ fluctuations. Using
Eq.(\ref{dispa}) we find
\beq
I\sim\frac{(CT)^{3/2}}{(2\pi)^{3/2}}\zeta(3/2)
\eeq
for $T\leq C$. However, as mentioned above Eq.~(\ref{dispa}) is no longer good
for $|{\bf p}|<\mu_K\sin\langle\theta\rangle$. We can estimate the
integral for this region
\beq
I_\delta\equiv \int_0^{\mu_K\sin\langle\theta\rangle}
\frac{p^2dp}{2\pi^2}f_B(E_-({\bf p}))
\sim \frac{C\mu_K\sin\langle\theta\rangle}{\pi^2\beta}.
\eeq
Therefore, the approximation (\ref{dispa}) is
verified,  if $I_\delta/I\ll 1$, which requires the following
condition for $c_3$ or $\mu_K\sin\langle\theta\rangle$ 
as the function of density and temperature,
\beq
c_3\ll \frac{C\pi}{16}T\zeta^2(3/2)\quad 
{\rm or} \quad 
\mu_K\sin\langle\theta\rangle\ll \frac{1}{4}(2\pi CT)^{1/2}\zeta(3/2).
\label{con}
\eeq
We can expect
Eq.~(\ref{con}) for relevant densities and temperatures \cite{ty}, except
very low temperature, where thermal effects should be trivially unimportant. 

Secondly, we can also 
expect the smallness of $c_3$ qualitatively: 
in the limit, $\langle\theta\rangle\rightarrow
0$ or $\mu_K\rightarrow 0$, $c_3$ gives no contribution. We also know
that $\langle\theta\rangle$ and $\mu_K$ are inversely proportional to each
other \cite{lee}. Hence we might expect that the $c_3$ term
becomes relatively small. 



\subsection{Thermodynamic relations}
The effective thermodynamic potential
$\Omega_{chiral}=-T\ln Z_{chiral}$
reads from Eq.~(\ref{zk}),
\beq
\Omega_{chiral}=\Omega_c+\Omega_{sc}+\Omega_K+\Omega_N,
\label{omegachiral}
\eeq
where $\Omega_c$ is the classical kaon contribution,
\beq
\Omega_c=V[-f^2m_K^2(\cos\langle\theta\rangle-1)
-1/2\cdot\mu_K^2f^2\sin^2\langle\theta\rangle],
\eeq
$\Omega_{sc}$ the subtraction term  to avoid the double-counting of
the interaction terms within the Hartree approximation,
\beq
\Omega_{sc}=T{\rm Tr}[G_H^{-1}\langle\delta G\rangle]
=V\left[\sigma\cos\langle\theta\rangle n_K^1
+bn_K^2\right].
\eeq
The kaon contribution, $\Omega_K=-T\ln Z_K^{eff}$,  can be written as,
\beq
\Omega_K=\Omega_K^{ZP}+\Omega_K^{th},
\eeq
where the zero-point energy (ZP) contribution $\Omega_K^{ZP}$ and 
the thermal one
$\Omega_K^{th}$ are given as follows;
\beq
\Omega_K^{ZP}=V\int \frac{d^3 p}{(2\pi)^3}\frac{1}{2}
[E_+({\bf p})+E_-({\bf p})]
\label{zp}
\eeq
and 
\beq
\Omega_K^{th}=TV\int \frac{d^3 p}{(2\pi)^3}
\ln(1-e^{-\beta E_+({\bf p})})(1-e^{-\beta E_-({\bf p})}).
\label{zkth}
\eeq



The thermodynamic potential 
for nucleons, $\Omega_N=-T\ln Z_N$, can be also written as follows:
\beq
\Omega_N=-T{\rm Tr}\ln G_H=\Omega_N^{ZP}+\Omega_N^{th}, 
\eeq
where $\Omega_N^{ZP}$ denotes the ZP contribution,
\beq
\Omega_N^{ZP}=-2V\sum_{n,p}\int\frac{d^3
p}{(2\pi)^3}\left[\frac{1}{2}(\epsilon_i+\epsilon_{\bar i})\right],
\label{zpn}
\eeq
and $\Omega_N^{th}$ the thermal contribution,
\beq
\Omega_N^{th}=-2TV\sum_{n,p}\int\frac{d^3p}{(2\pi)^3}
\left[\ln(1+e^{-\beta(\epsilon_i-\mu_i)})
+\ln(1+e^{-\beta(\epsilon_{\bar i}+\mu_i)})\right].
\eeq
Here, $\epsilon_i(\epsilon_{\bar i})$, $i=p,n$ are the single particle 
energies of the nucleons (anti-nucleons),
\begin{eqnarray}
\epsilon_{p,\bar p}&=&\mp\frac{1}{2f^2}n_K^2
\mp\mu_K(1-\cos\langle\theta\rangle)+E_p^*,\nonumber\\
\epsilon_{n,\bar n}&=&\mp\frac{1}{4f^2}n_K^2
\mp\frac{\mu_K}{2}(1-\cos\langle\theta\rangle)+E_n^*,
\end{eqnarray}
with the kinetic energy, $E_i^*=\sqrt{p^2+M^{*2}_i}$.

Using the thermodynamic
relations,
\beq
S_{chiral}=-\frac{\partial\Omega_{chiral}}{\partial T}, \quad
Q_i=-\frac{\partial\Omega_{chiral}}{\partial\mu_i}, \quad
E_{chiral}=\Omega_{chiral}+TS_{chiral}+\sum_i\mu_iQ_i,
\eeq
we  can find charge, entropy and internal energy. We easily see
that these quantities become too complicated to be tractable if we use 
the exact dispersion relation in Eq.~(\ref{disp})
; e.g. consider the kaonic charge 
given by way
of the relation, $Q_K=-\partial\Omega_{chiral}/\partial\mu_K$,
which results in a complicated form. Besides this, the quantities 
,$n_K^1$ or $n_K^2$, which can be written by the use of the
thermal Green's functions, has a complicated form as well 
(see Appendix B). Hence a useful approximation is desirable. We have
already discussed the pertinence of the approximation for the
dispersion relation  in Eq.~(\ref{dispa}) and we can see that this
approximation allows us to write down the thermodynamic quantities
in clear forms. 
For the kaonic charge it immediately gives
\beq
Q_K/V=\mu_Kf^2\sin^2\langle\theta\rangle
+\cos\langle\theta\rangle n_K
+(1+x)\sin^2(\langle\theta\rangle/2)\rho_B,
\label{qk}
\eeq
where $n_K$ is the number density of thermal kaons,  
\beq
n_K=\int\frac{d^3 p}{(2\pi)^3}
\left[f_B(E_-({\bf p}))-f_B(E_+({\bf p}))\right].
\label{nk}
\eeq
The resultant form (\ref{qk}) suggests that the charge of kaonic modes is
screened by the condensate; in fact, we can see that 
their effective charge are given by 
$\pm e_{eff}$ with $e_{eff}=\cos\langle\theta\rangle e$ by the use of the
effective Lagrangian (\ref{effl}). Accordingly 
the modified chemical potential $\tilde\mu_K=\cos\langle\theta\rangle
\mu_K$ gets the meaning of the chemical potential for the {\it number} 
of thermally excited kaons.

\section{Effects of zero-point fluctuations}
\label{eff}
\subsection{Renormalization}
In our formalism the thermodynamic potential (\ref{omegachiral}), as
well as the quantities $n_K^1$, $\sigma$, is
constructed by many one-loop bubbles, which include intrinsic 
divergences. 
Hence we must renormalize them to extract finite
contributions. Since the building units of these quantities 
are graphically one-loop bubbles of kaons
and nucleons, it suffices to renormalize these diagrams \cite{ser}. Then
all the diagrams with many loops have no more divergences. Since the
temperature dependent part never induces new divergence, we
renormalize the temperature independent part. Using the similar method 
proposed in ref. \cite{tho}, we can properly renormalize the thermodynamic
potential. However, it is to
be noted that we must carry out the renormalization program by
including not only the kaonic contributions but also the nucleonic one 
in this case.
The counterterms are introduced without spoiling the original symmetry 
structure of the Lagrangian; the symmetric terms are also invariant under
$\zeta$ and the symmetry-breaking terms should transform like Eq.~(2.23). 
Since we have already known the stucture of the effective Lagrangian,
given in Eqs.~(\ref{sk})-(\ref{sint}), it is easy to find the proper 
counterterms.
It is also to be noted that 
the vector-type contributions, $\delta{\cal L}$ induced by introducing
the chemical potentials as well as the original Tomozawa-Weinberg
term, never suffer from divergences (c.f. Eq.~(\ref{nk2}), see also
ref. \cite{ser}). Thereby it is almost clear the introduction of chemical
potentials never generates new divergences.  
In
the following we use the approximated formulae (\ref{dispa}) for the dispersion
relations of kaonic modes for simplicity.
 
First, we renormalize the self-energy terms of kaons and nucleons 
given in Eqs.~(\ref{effm}) and (\ref{effl}), respectively. The kaon effective
mass is given by
\beq
m_K^{*2}=m_K^2-\sigma.
\label{effkaonmass}
\eeq
The self-energy term $\sigma$ is given by the nucleon loop, 
$\sigma=f^{-2}{\rm Tr}\left
[ G_H^{-1}\gamma_0\Sigma_{KN}\right]=\sigma_{tree}+\sigma_{loop}$ 
with the inverse propagator
$G_H$ (\ref{gh}). We only 
consider the loop correction $\sigma_{loop}$ in the following.
To the one-loop order we don't care about the $n_K^1$
that is given by the kaon loop in the effective mass $M_i^*$
(\ref{effm}). Then 
the temperature independent part $\sigma_{loop}|_{T=0}$ reads
\begin{eqnarray}
\sigma_{loop}|_{T=0}&=&-4\sum_i\Sigma_{Ki}f^{-2}
\int\frac{d^3 k}{(2\pi)^3}\frac{M_i^*}{2E_i^*}\nonumber\\
&=&-4\sum_i M_i^*\Sigma_{Ki}f^{-2}\left[d_2-M_i^{*2}d_0
+\frac{1}{2}M_i^{*2}\ln\left(\frac{M_i^{*2}}{4\kappa^2}\right)\right]
\label{sigmaloop}
\end{eqnarray}
with the cut-off ($\Lambda$) regularization, where 
$d_i$'s are
the quadratically and logarithmically divergent integrals,
\begin{eqnarray}
d_2&=&\int_{k\le \Lambda}\frac{d^3 k}{(2\pi)^3}\frac{1}{2k}
=\frac{\Lambda^2}{8\pi^2},\nonumber\\
d_0&=&\int_{k\le \Lambda}\frac{d^3
k}{(2\pi)^3}\frac{1}{4k^3}=\frac{1}{8\pi^2}\ln\left(\frac{\Lambda}{\kappa}
\right),
\label{div}
\end{eqnarray}
with an arbitrary momentum-scale $\kappa$. To renormalize the
effective mass we redefine the kaon mass and the $KK$
interaction vertex. Then the counterterms should be given as 
\beq
\sigma_{ct}=4\sum_i M_i^*\Sigma_{Ki}f^{-2}[\alpha_2-M_i^{*2}\alpha_0].
\label{sigmact}
\eeq
The meaning of the counterterms $\alpha_i$ is graphically clear
(Fig.~4 ): $\alpha_2$ renormalizes the kaon mass and $\alpha_0$ the
$KK$ vertex at momentum ${\bf p}=0$. It is to be noted that the
wavefunction renormalization is not needed to the one-loop order.

\begin{figure}[h]
\epsfxsize=0.3\textwidth
\centerline{\epsffile{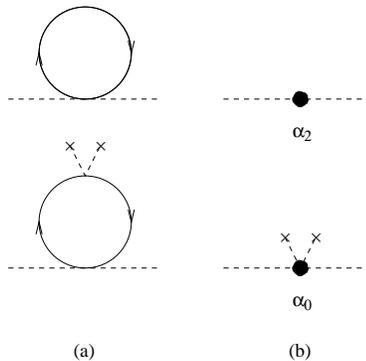}}
\vspace*{5mm}
\caption{Divergent diagrams (a) and the counterterms (b) to the one-loop
order for the kaon propagator (dashed line). Solid line denotes the nucleon
propagator without $n_K^1$ and the cross ($\times$) denotes the condensate 
$\langle\theta\rangle$. The black blob stands for the contribution from 
the counterterm.}
\end{figure}

Then they are determined by the following conditions,
\beq
(\sigma_{loop}|_{T=0}+\sigma_{ct})|_{vac}=0, \qquad 
\frac{\partial^2}{\partial\langle\theta\rangle^2}
(\sigma_{loop}|_{T=0}+\sigma_{ct})|_{vac}=0.
\eeq
From Eqs.~(\ref{sigmaloop}),(\ref{sigmact})we find 
\beq
\alpha_2=d_2-\frac{M^2}{16\pi^2}, \qquad 
\alpha_0=d_0+\frac{1}{8\pi^2}\left[\ln\left(\frac{2\kappa}{M}\right)-\frac{1}{2}\right], 
\label{counteralpha}
\eeq 
and consequently the finite contribution is given by
\beq
\sigma^{ren}_{loop}|_{T=0}=-\sum_iM_i^*\frac{\Sigma_{Ki}f^{-2}}{4\pi^2}
\left[M_i^{*2}\ln\left(\frac{M_i^{*2}}{M^2}\right)
+M^2-M_i^{*2}\right].
\label{rensigma}
\eeq
Thus the self-energy term $\sigma$ is renormalized to be $\sigma_{ren}$, 
$\sigma_{ren}=\sigma_{tree}+\sigma^{ren}_{loop}|_{T=0}$. Note that the 
dispersion relation is modified by the replacement of 
$\sigma$ by $\sigma_{ren}$, compared with the tree approximation where 
$\sigma$ is given by $\sigma=\sigma_{tree}$ \cite{ty}. However, we can 
see that the Goldstone-boson nature of $E_-$ is never spoiled by such
procedure (see Eq.~(\ref{fieldeq})).

Next, we renormalize the nucleon effective mass, $M^*_i (i=n,p)$, 
which is given by Eq.~(\ref{effm}).
It includes the kaon loop correction $n_K^1$
\footnote{Note that another kaon-loop contribution $n_K^2$ never
includes divergence (see Eq.~(\ref{nk2})) because it originates from 
the vector-type $KN$ interaction, while $n_K^1$ from the scalar-type
interaction.}
, the temperature independent part of which is given by
\begin{eqnarray}
n_K^1|_{T=0}&=&\int\frac{d^3p}{(2\pi)^3}\frac{1}{E_++E_-}\nonumber\\
&=&d_2-C^2d_0+\frac{1}{16\pi^2}C^2\ln\left(\frac{C^2}{4\kappa^2}\right),
\label{nk1t0}
\end{eqnarray}
(see Eq.~(\ref{nk1})).
Accordingly we introduce the counterterms, which renormalize the
nucleon mass and the $KN$ interaction vertex at momentum ${\bf
p}=0$ (see Fig.~5 ),
\beq
M_{ct}^i=f^{-2}\Sigma_{Ki}\cos\langle\theta\rangle[\beta_2-C^2\beta_0].
\label{mct}
\eeq 
Here $C^2$ is given by
\beq
C^2=(m_K^2-\sigma)\cos\langle\theta\rangle+b^2,
\eeq
but the nucleon loop in $\sigma$ is irrelevant up to the one-loop order
as well. It is to be noted that the overall factor 
$\cos\langle\theta\rangle$ in Eq.~(\ref{mct}) 
stems from the transformation $\zeta$ (see, 
Eqs.~(\ref{sk}) and (\ref{sint})).

\begin{figure}[h]
\epsfxsize=0.3\textwidth
\centerline{\epsffile{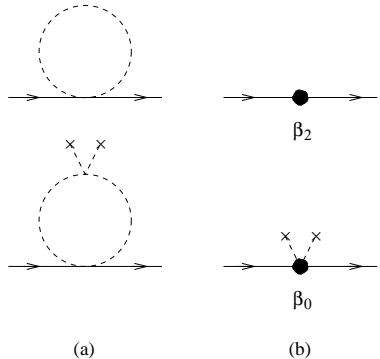}}
\vspace*{5mm}
\caption{Divergent diagrams (a) and the counterterms (b) to the one-
loop order for the nucleon propagator. Notations are the same as in Fig.~4 .}
\end{figure}

Then renormalization conditions are given as follows,
\beq
(M_i^*|_{T=0}+M_{ct}^i)|_{vac}=M, \qquad  
\frac{\partial^2}{\partial\langle\theta\rangle^2}(M_i^*|_{T=0}+M_{ct}^i)|_{vac}
=-\Sigma_{Ki}. 
\eeq
From Eqs.~(\ref{effm}),(\ref{nk1t0}), (\ref{mct}) we find 
\beq
\beta_2=d_2-\frac{m_K^2}{16\pi^2},  \qquad 
\beta_0=d_0+\frac{1}{8\pi^2}\left[\ln\left(\frac{2\kappa}{m_K}\right)
-\frac{1}{2}\right],
\label{counterbeta}
\eeq
and thereby the finite contribution to $n_K^1$ reads
\beq
n_K^{1, ren}|_{T=0}=\frac{1}{16\pi^2}
\left[m_K^2-C^2-C^2\ln\left(\frac{m_K^2}{C^2}\right)\right].
\label{rennk1}
\eeq
It is to be noted that the structure of (\ref{counterbeta}) is the
same as (\ref{counteralpha}).

Once $\sigma$ and $n_K^1$ are thus renormalized, we can use the
renormalized ones given by Eqs.~(\ref{rensigma}) and (\ref{rennk1}) in 
the thermodynamic potential (\ref{omegachiral})
instead of the unrenormalized ones.
Although the thermodynamic potential (\ref{omegachiral}) still includes the
divergences through the ZP contributions in Eq.~(\ref{zp}) and 
Eq.~(\ref{zpn}),  the total ZP
contribution $\Omega_{ZP}=\Omega_K^{ZP}+\Omega_N^{ZP}$ should be renormalized 
by the use of the counterterms
$\alpha_i$ and $\beta_i$ without introduction of another counterterms 
(see Fig.~6 ). 

\begin{figure}[h]
\epsfxsize=0.3\textwidth
\centerline{\epsffile{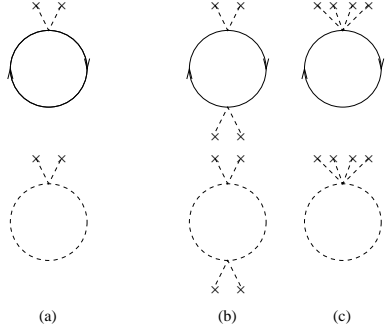}}
\vspace*{5mm}
\caption{Divergent diagrams to the one-loop order 
for the thermodynamic potential. Notations 
are the same as in Fig.~4. The diagrams in (a) are quadratically
divergent and renormalized by the counterterms given in Figs.~4 and ~5
, while the diagrams given in (b) and (c), which are renormalized by
redefining $KK$ interaction vertex at momentum ${\bf p}=0$, 
give the quadratic and logarithmic 
divergences, respectively; the counterterms for (b) and (c) are
already given in Figs.~4 and ~5 .}
\end{figure}

\noindent Thus we find the renormalized ZP contribution, 
$\Omega_{ZP}^{ren}$,
\beq
\Omega_{ZP}^{ren}=\Omega_{ZP,K}^{ren}+\Omega_{ZP,N}^{ren},
\label{zpren}
\eeq
with the kaon-loop contribution,
\beq
\Omega_{ZP,K}^{ren}/V=\frac{1}{64\pi^2}\left[(C^2-m_K^2)(m_K^2-3C^2)+
2C^4\ln\left(\frac{C^2}{m_K^2}\right)\right],
\label{zprenk}
\eeq
and the nucleon-loop one,
\beq
\Omega_{ZP,N}^{ren}/V=-\sum_{i=n,p}\frac{1}{32\pi^2}
\left[(M_i^{*2}-M^2)(M^2-3M_i^{*2})+
2M_i^{*4}\ln\left(\frac{M_i^{*2}}{M^2}\right)\right].
\label{zprenn}
\eeq
Here we have subtracted the vacuum energy which is quartically
divergent, 
\beq
\Omega_{vac}/V=-3d_4-\frac{M^4}{16\pi^2}+\frac{m_K^4}{64\pi^2}
\eeq
with the divergent integral,
\beq
d_4=\int_{k\le \Lambda}\frac{d^3 k}{(2\pi)^3}k
=\frac{\Lambda^4}{8\pi^2}.
\eeq
Then we can easily verify the relations,
\beq
\frac{\partial(\Omega_{ZP,K}^{ren}/V)}{\partial C^2}=n_K^{1,ren}|_{T=0},
\qquad
\sum_{i=n,p}f^{-2}\Sigma_{Ki}\frac{\partial(\Omega_{ZP,N}^{ren}/V)}{\partial
M_i^*}=\sigma^{ren}_{loop}|_{T=0}.
\label{renrel}
\eeq
It is to be noted that the magnitude of ZP contributions (\ref{zprenk}) and
(\ref{zprenn}) are bounded from above, 
\beq
|\Omega_{ZP,K}^{ren}/V|< 12{\rm MeV\cdot fm^{-3}} \quad{\rm and}\quad 
|\Omega_{ZP,N}^{ren}/V|< 620{\rm MeV\cdot fm^{-3}}.
\eeq
Comparing the both limit values, we can see that the effect of 
the zero-point fluctuation 
by nucleons should be more pronounced than that by kaons.

In the infinite-mass limit for nucleons, $M\rightarrow\infty$, there
is only left the kaon contribution and 
$\Omega_{ZP}^{ren}$ is reduced to a simple form \cite{ty},
$\Omega_{ZP}^{ren}\rightarrow \Omega_{ZP,K}^{ren}$.
It may be worth noting the relations, $\beta_2-d_2=f_2$ and 
$(\beta_0-d_0)+(\beta_2-d_2)/m_K^2=f_0$ in terms of $f_i$ given in 
ref.~\cite{tho}, which is graphically understood in Fig.~6 . 

Finally, we give a remark about the use of the
approximated formulae (\ref{dispa}) for the dispersion relations of the
kaonic modes. As has been already mentioned in \ref{the}, the difference
between (\ref{dispa}) and the exact formulae (\ref{disp}) converges to 
zero for high-momentum limit, so that the structure of the
ultraviolet divergences is the same to each other. Therefore, even if
we use the exact formulae, we find the same counterterms in 
Eq.~(\ref{counterbeta}).

\subsection{Nucleon-loop contribution}

We investigate the effects of zero-point fluctuations in neutron-star 
matter at $T=0$, 
since the negative-sea effect is dominant there.
At $T=0$ the thermodynamic potential is reduced into a
simple form,
\beq
\Omega_{chiral}|_{T=0}=\Omega_c+\Omega_{sc}|_{T=0}+\Omega_N|_{T=0}
+\Omega_{ZP}^{ren},
\eeq
where $\Omega_{sc}|_{T=0}
=V[\sigma_{ren}n_K^{1,ren}|_{T=0}\cos\langle\theta\rangle]$. The
nucleon contribution $\Omega_N|_{T=0}$ can be written as 
\begin{eqnarray}
\Omega_N|_{T=0}&\simeq&
V\left[3/5\epsilon_F^n(1-x)+3/5\epsilon_F^px
-\mu_n(1-x)-\mu_px\right]\rho_B\nonumber\\
&-&(2b\mu_K+\sigma_{tree})f^2(1-\cos\langle\theta\rangle)
\end{eqnarray}
with the baryon
number density $\rho_B$, the proton mixing
ratio $x=\rho_p/\rho_B$, and Fermi energies for nucleons, 
$\epsilon_F^n=\left(3\pi^2\rho_B(1-x)\right)^{2/3}/2M$, 
$\epsilon_F^p=(3\pi^2\rho_Bx)^{2/3}/2M$, 
in the nonrelativistic approximation. Since we have already known that
the kaon contribution  $\Omega_{ZP,K}^{ren}$ gives only a tiny effect 
\cite{tho,ty}, we, hereafter,  concentrate on the nucleon one 
$\Omega_{ZP,N}^{ren}$ by simply dropping kaon loops. 
After the calculation we will confirm it as a consistency check.

It is well-known that 
the nuclear symmetry energy plays an important role for the
ground-state properties of the condensed phase. Hence we take it into
account to get a realistic result \cite{ty}.
Since we have already included 
the kinetic energy for nucleons, it suffices to consider only the potential
energy contribution. Following Prakash et al. \cite{prak} we
effectively introduce a symmetry energy contribution,
\beq
\Omega_N^{symm}=V\rho_B(1-2x)^2S^{pot}(u),
\eeq  
with the relative density, $u=\rho_B/\rho_0$. 
The coefficient $S^{pot}(u)$ is given as
$
S^{pot}(u)=(S_0-(2^{2/3}-1)(3/5)\epsilon_F^0)F(u)
$
with the constraint $F(1)=1$, 
to reproduce the empirical symmetry energy $S_0\simeq 30$MeV, where 
$\epsilon_F^0$ is the Fermi energy at $\rho_0$, 
$\epsilon_F^0=(3\pi^2\rho_0/2)^{2/3}/2M$ , and $F(u)$ is a simulated
function, for which  
we take here $F(u)=u$ for simplicity.

Then the total thermodynamic potential, we consider here, is given by further 
adding the leptonic one $\Omega_l|_{T=0}$,
\beq
\Omega_{tot}=\Omega_{chiral}|_{T=0}+\Omega^{symm}_N+\Omega_l|_{T=0},
\label{omegatotal}
\eeq
where $\Omega_l|_{T=0}$ can be written in the standard noninteracting form,
\beq
\Omega_l|_{T=0}=V\left[-\frac{\mu_K^4}{12\pi^2}
+\theta(|\mu_K|-m_\mu)\left\{\frac{m_\mu^4}{8\pi^2}[(2t^2+1)t\sqrt{t^2+1}
-\ln(t+\sqrt{t^2+1})]-\frac{m_\mu^3|\mu_K|t^3}{3\pi^2}\right\}\right],
\eeq
with $t=\sqrt{\mu_K^2-m_\mu^2}/m_\mu$. Here the first term is due to
electrons and the second term due to muons, and we have used the relation 
$\mu_l=\mu_K$ for leptons.

The parameters $\langle\theta\rangle$, $x$ and $\mu_K$ are determined 
by the equilibrium conditions and charge neutrality of the ground
state \cite{lee,tat,tho}. First, we get 
\beq
f^2\sin\langle\theta
\rangle(m_K^2-\sigma_{tree}-2\mu_Kb-\mu_K^2\cos\langle\theta\rangle)
+\frac{\partial(\Omega_{ZP,N}^{ren}/V)}
{\partial\langle\theta\rangle}=0
\label{cond1}
\eeq
from the condition
$\partial\Omega_{tot}/\partial\langle\theta\rangle=0$. Recalling the
relation in Eq.~(\ref{renrel}), we find it is further written as
\beq
f^2\sin\langle\theta
\rangle(m_K^{*2}-2\mu_Kb-\mu_K^2\cos\langle\theta\rangle)=0,
\label{fieldeq}
\eeq
with the effective kaon mass $m_K^{*2}=m_K^2-\sigma_{ren}$. Then it is 
to be noted that the critical density is unchanged even if the
zero-point fluctuation $\Omega_{ZP,N}^{ren}$ is included, since
$\sigma_{ren}$ is reduced to $\sigma_{tree}$ as $\langle\theta\rangle
\rightarrow 0$. It is also to be noted that Eq.~(\ref{fieldeq})
ensures the Goldstone-boson nature of $K^-$ mode (see Eqs.~(\ref{disp}), 
(\ref{dispa})), even if the ZP contribution is included. Secondly, we get
\beq
\epsilon_F^p-\epsilon_F^n-4S^{pot}(u)(1-2x)
+\frac{1-\cos\langle\theta\rangle}{2}
\{2(\Sigma_{Kn}-\Sigma_{Kp})-\mu_K\}+\mu_K=0.
\label{cond2}
\eeq
from the condition $\partial\Omega_{tot}/\partial x=0$, which is
equivalent with the chemical equilibrium condition among kaons and
nucleons, $\mu_n-\mu_p=\mu_K$.
Finally, charge neutrality demands the relation, 
\beq
Q_K+Q_l=Q_p(=xV\rho_B),
\label{cha}
\eeq
which implies $\partial\Omega_{tot}/\partial\mu_K=0$ by definition,
where the lepton charge is given by 
\beq
Q_l/V=\frac{\mu_K^3}{3\pi^2}+\theta(|\mu_K|-m_\mu)\frac{\mu_K}{|\mu_K|}
\frac{m_\mu^3t^3}{3\pi^2},
\eeq
and the kaonic charge,
\beq
Q_K/V=\mu_Kf^2\sin^2\langle\theta\rangle
+(1+x)\sin^2(\langle\theta\rangle/2)\rho_B.
\eeq  
Then the energy is given by way of the thermodynamic relation, 
\beq
E_{tot}=\Omega_{tot}+\sum_{n,p,K,l}\mu_iQ_i.
\eeq
Note that these equations are the same as the previous ones in the
heavy-nucleon limit \cite{ty} if we simply put $\Omega_{ZP,N}^{ren}=0$.

In Table I and II we present the optimum values of $\mu_K, x,$ and 
$\langle\theta\rangle$ for given densities, in comparison with the
classical values. We use the values, 
$a_1m_s=-67$ MeV, 
$a_2m_s=134$ MeV and $a_3m_s=-134(-222)$ MeV in Table I(II). 
\footnote{Recent lattice simulations suggest larger values for $KN$
sigma term \cite{fuk}.}
\clearpage

\begin{table}[h]
 \caption{Optimum values of the parameters for the tree case and the
one-loop case. We use the value, $a_3m_s = -134$MeV.}
 \begin{center}
  \begin{tabular}{rrrrrrr} \hline
\multicolumn{1}{c}{} &
\multicolumn{3}{c}{tree} &
\multicolumn{3}{c}{tree + loop} \\
\multicolumn{1}{c}{$u$} & 
\multicolumn{1}{c}{$\langle \theta \rangle [{\rm deg.}]$} & 
\multicolumn{1}{c}{$\mu_K$ [{\rm MeV}]} & 
\multicolumn{1}{c}{$x$}	& 
\multicolumn{1}{c}{$\langle \theta \rangle [{\rm deg.}]$} & 
\multicolumn{1}{c}{$\mu_K$ [{\rm MeV}]} & 
\multicolumn{1}{c}{$x$}	\\ \hline
4.19 & 0.00 &  255.8 & 0.193 & 0.00 &  255.8 & 0.193 \\
4.40 & 18.85 & 244.0 & 0.231 & 18.74 & 244.2 & 0.231 \\
4.80 & 31.97 & 221.1 & 0.293 & 31.49 & 222.6 & 0.290 \\
5.20 & 40.85 & 197.7 & 0.343 & 39.91 & 201.6 & 0.338 \\
5.60 & 47.62 & 174.4 & 0.384 & 46.26 & 181.3 & 0.376 \\
6.00 & 52.92 & 151.9 & 0.417 & 51.26 & 161.9 & 0.407 \\
6.40 & 57.09 & 130.6 & 0.442 & 55.25 & 143.5 & 0.431 \\
6.80 & 60.36 & 110.8 & 0.462 & 58.46 & 126.0 & 0.450 \\
7.20 & 62.91 & 92.6 & 0.477 & 61.05 & 109.7 & 0.465 \\ \hline
  \end{tabular}
 \end{center}
\end{table}%
\vskip 2cm

\begin{table}
 \caption{Optimum values of the parameters for $a_3m_s = -222$MeV.}
 \begin{center}
  \begin{tabular}{rrrrrrr} \hline
\multicolumn{1}{c}{} &
\multicolumn{3}{c}{tree} &
\multicolumn{3}{c}{tree + loop} \\
\multicolumn{1}{c}{$u$} & 
\multicolumn{1}{c}{$\langle \theta \rangle [{\rm deg.}]$} & 
\multicolumn{1}{c}{$\mu_K [{\rm MeV}]$} & 
\multicolumn{1}{c}{$x$}	& 
\multicolumn{1}{c}{$\langle \theta \rangle [{\rm deg.}]$} & 
\multicolumn{1}{c}{$\mu_K$ [{\rm MeV}]} & 
\multicolumn{1}{c}{$x$}	\\ \hline
3.08 & 0.00 & 218.8 & 0.156 & 0.00 & 218.8 & 0.156 \\
3.20 & 18.28 & 206.4 & 0.198 & 17.42 & 207.9 & 0.195 \\
3.60 & 39.75 & 160.0 & 0.326 & 33.60 & 181.3 & 0.286 \\
4.00 & 53.71 & 109.2 & 0.425 & 42.20 & 161.8 & 0.344 \\
4.40 & 63.74 & 59.1 & 0.494 & 48.33 & 144.7 & 0.386 \\
4.80 & 71.05 & 14.4 & 0.538 & 53.12 & 128.5 & 0.419 \\
5.20 & 76.13 & -23.3 & 0.565 & 57.04 & 112.3 & 0.444 \\
5.60 & 79.67 & -54.9 & 0.581 & 60.30 & 95.8 & 0.465 \\
6.00 & 82.19 & -81.5 & 0.591 & 63.08 & 79.2 & 0.482 \\ \hline
  \end{tabular}
 \end{center}
\end{table}%

Then we can observe 
a sizable effect of the zero-point fluctuation by nucleons; it
shifts the optimum values of $x$ and $\langle\theta\rangle$ downward,
while $\mu_K$ upward, comapared with the classical values. The
effect becomes prominent as the value of the $KN$ sigma term is
increased, since the reduction rate of the nucleon effective mass
becomes high. 
As mentioned above it never affects the critical
density, while it suppresses the growth of the condensate.

In Fig.7 the energy difference between the condensed phase and normal 
one is dipicted as a function of density. We can see  that the energy 
gain is also
reduced along with the suppression of the condensate; the zero-point
fluctuation of nucleons reduces the energy gain by shifting the
optimum values 
of parameters, while the resultant ZP contribution is small in magnitude.
We also estimate the magnitude of the kaon loops by the use of the
values given in Table I, II to be tiny for a consistency check. 

\begin{figure}[h]
\epsfxsize=0.6\textwidth
\centerline{\epsffile{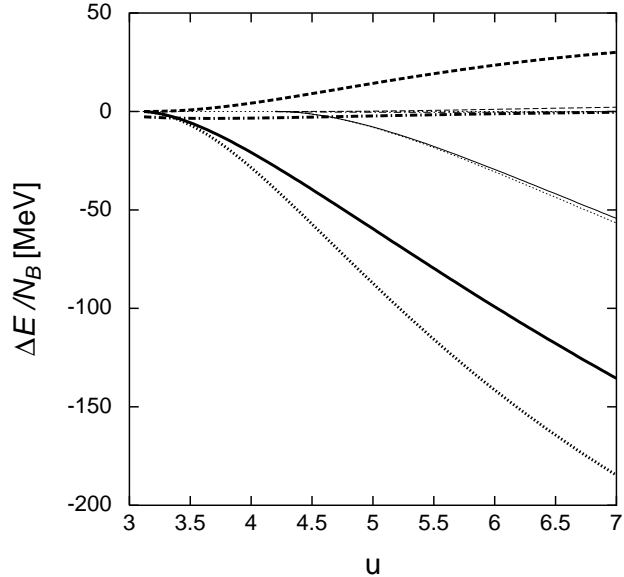}}
\caption{Energy differences per baryon, 
$\Delta E/N_B\equiv (E_{tot}-E_{tot}(\langle\theta\rangle=0))/N_B$,
and ZP contributions   
for $a_3m_s = -134$MeV (thin lines) and $a_3m_s = -222$MeV (thick lines). 
Solid lines show the results with the zero-point fluctuation by 
nucleon, while dotted lines those in the tree level. Dashed lines 
show the ZP contribution. Kaon ZP contributions are estimated 
by the use of optimum values given in Table I, II (dash-dotted lines).}
\end{figure}

We notice that these consequences originate from the replacement of
the $KN$ interaction term $\sigma_{tree}$ in the tree approximation by 
$\sigma_{ren}$ in the effective mass of kaons $m_K^*$; 
the zero-point fluctuation works to reduce the attractive effect. 
It may be worth noting that 
$m_K^{*2}$ stays to be in positive value and $\mu_K$ also becomes
positive, once the zero-point fluctuation is included. 
It may remind us of the relativistic effect by nucleons \cite{maru}, which
also gives rise to the same effect. Hence, if we further 
take into account the 
relativistic effect in our formalism, it may be milder than before.
Dispesion relations are also checked to be fairly good by comparing
the results given by Eqs.~(\ref{disp}) and (\ref{dispa}).

\section{Summary and concluding remarks}
\label{sum}
In this paper we have presented a formalism to treat fluctuations around 
the condensate within the framework of chiral symmetry. 
Our approach is based on the group theoretical
argument; Goldstone fields are regarded as coordinates on the
chiral $SU(3)_L\times SU(3)_R/SU(3)_V$ manifold and each state corresponding
to the condensed phase or the normal phase with fluctuations can be
represented as a chiral rotated state from the vacuum, which
corresponds to the fiducial point on the manifold. Hence the
fluctuations in the vicinity of the condensate are introduced by way of the
successive chiral transformations by $\zeta$ and $\eta$. This
procedure can be regarded as the introduction of the local coordinates 
around the condensed point on the chiral manifold. Since the chiral
transformation is the isometry, the geometry around the condensed
point is the same as in the vicinity of the vacuum, where we might use 
the flat curvature. Hence this method
has the advantage of avoiding the Lee-Yang term. 

More interestingly, this method corresponds to the separation of
the zero mode; since there should appear the Goldstone mode as a
result of the symmetry breaking of $V$-spin beyond the critical
density, we must treat the zero modes nonperturbatively even in the normal
phase. In the infinite-volume limit the condensate only contributes to 
the thermodynamical potential among zero modes in the condensed phase.
Therefore we have put $U$ into the form given in Eq.~(2.14) from the beginning.
It is worth mentioning that this method might have a wider applicability
for studying phase transitions, e.g. in finite volume, if system has
some symmetry. 

We have considered the thermodynamic potential up to the one-loop
diagrams of kaons and nucleons and 
the $KN$ interactions have been treated self-consistently within the
Hartree approximation; consequently it consists of infinite series of 
many one-loops.
Then the dispersion relation for kaon excitation, which is a 
fundamental object to examine the thermal properties of  
the condensed phase, can be obtained. There appear two modes:
Goldstone-like soft mode and very massive mode, which
correspond to $K^-$- and $K^+$- mesonic excitations,
respectively, in the condensed phase. Hence, the thermal loops due to
the soft mode play an important role. 
We have seen that the form of
the dispersion relation can be reduced to a simple one, if we can take 
an approximation, $c_3=0$. It should be
worth noting that once this approximation works, the expressions of
other quantities like charge, entropy, internal energy become very
simple.

The thermodynamic potential as well as the self-energy terms for kaons 
and nucleons contains some divergences. We have shown that these divergences 
are properly renormalized by way of the cut-off regularization; 
counterterms are introduced to redefine the 
masses of kaons and nucleons and the $KN$ and $KK$
interaction vertices. Once the self-energy terms are renormalized, we 
need not require more counterterms to renormalize the thermodynamic
potential.

We have discussed the effect of the zero-point fluctuation by nucleons. It
provides a sizable effect for the optimum values of the parameters, so 
that the energy gain is reduced from the classical values. This is
because the $KN$ sigma term remarkably reduces the nucleon effective
mass in the condensed phase, while the critical density is not
affected. On the other hand the contribution by the kaon loops is
still tiny.

It may be interesting to study thermal effects in the condensed phase,
which is closely related with the problem of protoneutron stars,
especially the effects of thermal loops of $n_K^{1, ren}, n_K^2$ ~\cite{kol}. 
In a subsequent paper~\cite{yas} we discuss the thermal properties 
of the kaon
condensed phase on the basis of the formalism given in this paper.
We shall see the phase diagram in the density-temperature plane or
equation of state for the kaon-condensed phase.

\vskip 1cm
\centerline{Acknowledgment}

We thank T.Muto for his interest in this work.
We also thank K. Matsuyanagi for useful discussions about the finite
volume effect on phase transition. This work was supported in 
part by the Japanese Grant-in-Aid sor
Scientific Research Fund of the Ministry of Education, Science and
Culture (08640369).

\newpage
\appendix
\section{Separation of zero modes}
Here we reconstruct our formalism from the viewpoint of separation of
the zero mode, following ref.\cite{gass}. To make our argument clear 
we first consider the finite 
volume $V=L^3$ and put it infinity at the end. 
Since we are interested in the kaon condensation, we restrict
ourselves to the kaon and nucleon degrees of freedom for simplicity. 
Considering the partition function (\ref{part}) and 
integrating out the baryonic degrees of freedom, we get the
partition function for $U$,
\beq
  Z_{chiral}=N'\int [dU]\exp[S^{eff}(U)],
\label{zu}
\eeq 
with the effective action $S^{eff}(U)=S(U)+{\rm Tr}\ln G(U)$, where $S(U)$
stands for the pure mesonic action and the second term stems from the
integration over the baryon field. The capital letter for trace is
used for the Dirac operator $G(U)$ of infinite dimension over 
space-imaginary-time.
Substituting $U=\exp(2i\phi/f)$ with $\phi=T_a\phi_a$ in Eq.~(\ref{zu}),
the action has the following structure, 
\beq
S^{eff}(U)=\bar S_K(K)+{\rm Tr}\ln \bar G,
\label{seffu}
\eeq
with 
\beq
\bar S_K=\int_0^\beta d\tau\int d^3x\left [-\partial_\mu K^+\partial_\mu K^-
-(m_K^2-\mu_K^2) K^+ K^- -\mu_K K^+
\stackrel{\leftrightarrow}{\partial_\tau} K^-+O(|K|^4)\right].
\eeq
The matrix $\bar G$ reads
\beq
\bar G=\bar G_0+\delta\bar G
\eeq
with the free part 
\beq
\bar G_0=-\frac{\partial}{\partial\tau}
+i\gamma_0\mbox{\boldmath$\gamma$}\cdot\mbox{\boldmath$\nabla$}-\gamma_0 M
+\mu_N
\eeq
and the interaction part
\beq
\delta \bar G=\frac{\Sigma_{KN}}{f^2}\gamma_0K^+K^-
-\frac{1}{f^2}\frac{3+\tau_3}{8}
\left(K^+\stackrel{\leftrightarrow}{\partial_\tau} K^-
-2\mu_K K^+K^-\right)+O(|K|^4).
\label{delg}
\eeq
For the purpose of the Hartree approximation, we define 
$\bar G_H$ as
\beq
\bar G_H=\bar G_0+\langle\delta\bar G\rangle, 
\label{delgh}
\eeq
by taking the thermal average for the kaon field. Then the effective
action of the nucleonic part can be written as 
\beq
{\rm Tr}\ln \bar G={\rm Tr}\ln \bar G_H+{\rm Tr}[{\bar G_H}^{-1}\delta G]
-{\rm Tr}[\bar G_H^{-1}\langle\delta \bar G\rangle]+O(|K|^4).
\eeq
Thus effective action $S^{eff}$ can be written as 
\beq
S^{eff}=\bar S_K+{\rm Tr}\ln\bar G_H+{\rm Tr}[{\bar G_H}^{-1}\delta G]
-{\rm Tr}[{\bar G_H}^{-1}\langle\delta G\rangle]+O(|K|^4).
\label{effective}
\eeq
Substituting Eqs.~(\ref{delg}) into
Eq.~(\ref{effective}), we find
\begin{eqnarray}
S^{eff}&=&{\rm Tr}\ln\bar G_H-{\rm Tr}[{\bar G_H}^{-1}\langle\delta
G\rangle]\nonumber\\
&+&\int_0^\beta d\tau\int d^3x\Bigl [-\partial_\mu
K^+\partial_\mu K^-
-M_K^2K^+ K^--(\bar b+\mu_K)K^+
\stackrel{\leftrightarrow}{\partial_\tau} K^-+O(|K|^4)
\Bigr ],
\label{aseff}
\end{eqnarray}
where $M_K^2$ is the ``mass'' term for the kaon field, 
$M_K^2\equiv m_K^2-\Pi_{KN}$, 
with the self-energy term, $\Pi_{KN}=-\bar \sigma-2\mu_K\bar b$. Here
$\bar\sigma$  and $\bar b$ stand for 
$\bar\sigma=f^{-2}{\rm Tr}[\gamma_0\Sigma_{KN}{\bar G_H}^{-1}]$ and 
$\bar b=f^{-2}{\rm Tr}[(3+\tau_3)/8{\bar G_H}^{-1}]$, respectively. 
It is to be noted that the ``mass'' term  
should include the
contributions from the meson-baryon interactions ($\Pi_{KN}$) and the chemical 
potential besides the genuine mass of kaons, $m_K^2$.

To perform the integral in Eq.~(\ref{zu}) we expand 
the Goldstone field in terms of periodic plane waves like Eq.~(\ref{period}),
\beq
\phi_a=\sqrt{\frac{\beta}{V}}\sum_n
\exp[2\pi i({\bf n}\cdot {\bf r}/L+n_4\tau T)]\phi_a^n.
\label{zero}
\eeq
Inserting Eq.~(\ref{zero}) into Eq.~(\ref{aseff}), we can see that 
the magnitude of $\phi_a^n$ is at most $O(T/\sqrt{M_K^2+L^{-2}})$. Hence 
if $T\ll M_K$, the Gaussian integral should give the
dominant contribution to the partition function. 

However, if the
phase transition occurs and  the ``mass'' term $M_K^2$ tends to 
be vanished at the critical
point, the standard chiral perturbation fails, because there is no
quadratic term for the zero-mode $(n_\mu=0)$; in fact, the ``mass'' term 
takes negative values before the phase 
transition, while it is vanished at the critical density.
 Hence we need to reorder 
the perturbation series by treating the zero mode nonperturbatively to 
get a consistent formulation valid even at the critical point. 

Following Gasser and Leutwyler we treat the zero-modes as the
collective variables \cite{gass},
\beq
 U= v U_f v
\label{vuv}
\eeq
with $U_f=\exp[2i(T_4\eta_4+T_5\eta_5)/f]$ and the $U(1)$ matrix 
$v=\exp(i\theta_0T_4)$. 
$v$ is the {\it constant} unitary matrix representing the zero mode 
and $\eta_a$ sums up the nonzero modes,
\beq
\eta_a=\sqrt{\frac{\beta}{V}}\sum_{n\neq 0}
\exp[2\pi i({\bf n}\cdot {\bf r}/L+n_4\tau T)]\phi_a^n.
\label{nonzero}
\eeq
After integrating out with respect to the nonzero modes, we
find the integral over the zero mode ($U(1)\subset SU(3)$)
\footnote{In the general case the zero modes cover the $SU(3)$
manifold, and the partition function is given by the group integral
over $SU(3)$ \cite{gas}.}
. Substituting Eq.~(\ref{vuv}) into Eq.~(\ref{seffu}), we can see that
the form of the effective action is the same with Eq.~(\ref{zch}) with
$\theta_0$ instead of $\langle\theta\rangle$. Hence the integral over
the zero mode becomes 
\beq
Z_{chiral}=\int d\theta_0
\exp[-\bar\Omega_{chiral}(\theta_0)/T],
\label{steep}
\eeq
where $\bar\Omega_{chiral}(\theta_0)$ is given by Eq.~(\ref{omegachiral})
with replacing the thermal average $\langle\theta\rangle$ by $\theta_0$.
Separating the trivial volume
dependence from the thermodynamic potentials, we write  Eq.~(\ref{steep})
as
\beq
Z_{chiral}=\int d\theta_0\exp[-V\beta\bar\omega],
\label{steepest}
\eeq
with $\bar\omega\equiv \bar\Omega_{chiral}(\theta_0)/V$. In 
the large volume limit or the low temperature limit we can apply the
steepest descent method to evaluate the integral in Eq.~(\ref{steepest}):
\beq
Z_{chiral}\simeq \exp[-V\beta\bar\omega(\theta_0^m)]
\sqrt{\frac{2\pi}{V\beta{\bar\omega}''}},
\label{decent}
\eeq
where $\omega''$ stands for 
$\partial^2\bar\omega/\partial\theta_0^2|_{\theta_0=\theta_0^m}$ and 
the saddle point $\theta_0^m$ is determined by the equation,
\beq
\partial\bar\omega(\theta_0)/\partial\theta_0|_{\theta_0=\theta_0^m}=0.
\label{extrem}
\eeq
This is nothing but an extremum condition for the thermodynamic
potential with respect to the order parameter.
Hence $\theta_0^m$ is equivalent with the thermal average of $\theta$, 
$\theta_0^m\equiv \langle\theta\rangle$. Accordingly the zero-mode
matrix $v_m$ is reduced to $\zeta$. In conclusion, when some degrees of
freedom should condense, it is enough to separate them from the
beginning in the form Eq.~(\ref{vuv}) in the 
limit, $V\rightarrow \infty$.

Here it is also interesting to observe that there is a logarithmic
singularity in the 
thermodynamic potential at the critical point as 
long as $V\neq \infty$.

\newpage
\section{Classical approximation -- (model-independent)}

In the classical approximation, $\phi_a=0$, the operator $u$ becomes
trivial, 
\beq
U_f=u=1.
\eeq
Furthermore, replacing the bilinear operator of the nucleon field by
its expectation value, e.g., 
$\bar\psi\psi\rightarrow\langle\bar\psi\psi\rangle$,
then we get
\begin{eqnarray}
\delta{\cal L}&=&\frac{\mu_K}{2}{\rm tr}
\{B^\dagger[(\zeta^\dagger[T_{em}, \zeta]+\zeta[T_{em}, \zeta^\dagger]), B]\}
+\frac{f^2\mu_K^2}{4}{\rm tr}\{[T_{em}, \zeta^2][T_{em},\zeta^{\dagger 2}]\}
\nonumber\\
&=&\mu_K(\cos\langle\theta\rangle-1)(1+x)\rho_B-
1/2\cdot\mu_K^2f^2\sin^2\langle\theta\rangle,
\end{eqnarray}
which stems from the kinetic terms of Goldstone bosons,
and thereby model-independent \cite{tat}.

The symmetry breaking part is also transformed, 
\begin{eqnarray}
{\cal L}_{SB}(U_K=\zeta^2, B)&=&{\cal L}_{SB}(1, B)+
(\cos\langle\theta\rangle-1)\langle\Sigma_K(1, B)\rangle   \nonumber\\
&=&(\cos\langle\theta\rangle-1)(-f^2m_K^2+\Sigma_{Kp}\rho_{p}
+\Sigma_{Kn}\rho_{n})+{\cal L}_{SB}(1, B),
\end{eqnarray}
according to the octet dominance hypothesis (see,
Eq.~(\ref{sbtrans})), where we have used the nonrelativistic
approximation for nucleons. 
Therefore the result is also
model independent once we employ this hypothesis \cite{tat}. 
Thus we can recover the previous result: for the energy difference
between kaon condensed phase and normal one,
\beq
\delta E/V=E_{cl}/V+\mu_KQ_{cl}^K/V,
\eeq
where the ``classical'' energy $E_{cl}$, which is nothing but the
classical thermodynamic potential, is given by
\beq
E_{cl}/V=(\cos\langle\theta\rangle-1)
[\mu_K(1+x)\rho_B/2-f^2m_K^2+\Sigma_{Kp}\rho_{p}
+\Sigma_{Kn}\rho_{n})]
-1/2\cdot\mu_K^2f^2\sin^2\langle\theta\rangle,
\eeq
for $T=0$. The charge of the condensed kaons, $Q_{cl}^K$ is given by
\beq
Q_{cl}^K/V=\mu_K^2f^2\sin^2\langle\theta\rangle
+\mu_K(1+x)\rho_B(1-\cos\langle\theta\rangle)/2.
\eeq
It is to be noted that this result has been obtained without recourse
to the definite Lagrangian. 

The optimum values of the parameters, $\langle\theta\rangle, x,
\mu_K$, are determined by the extremum conditions for $E_{cl}$, 
\beq
\partial E_{cl}/\partial\langle\theta\rangle=0,\quad 
\partial E_{cl}/\partial x=0, \quad
\partial E_{cl}/\partial\mu_K=0.
\eeq
In particular, we find the field equation for the condensate from the
first equation,
\beq
f^2\sin\langle\theta\rangle(m_K^{*2}-2\mu_Kb-\mu_K^2\cos\langle\theta\rangle)
=0
\eeq
with $m_K^{*2}=m_K^2-\sigma_{tree}$, 
which implies $c_1({\bf p}=0)=0$ for $\langle\theta\rangle\neq 0$.


\newpage
\section{Kaon Green's functions}

The Green's functions of kaons can be extracted from the matrix
$D^{eff}$. Define the inverse- propagator matrix 
${\cal D}^{-1}\equiv \beta^{-2} D^{eff}$: 
${\cal D}_{11}^{-1}=\omega_n^2+c_1+2c_3$, 
${\cal D}_{22}^{-1}=\omega_n^2+c_1$, 
and ${\cal D}_{12}^{-1}=-{\cal D}_{21}^{-1}=-2c_2\omega_n$. 
Then the propagators are given by
\beq
\langle \phi_4\phi_4\rangle
={\cal D}_{11}={\cal D}^{-1}_{22}|{\cal D}^{-1}|^{-1}
\eeq
and
\beq
\langle \phi_5\phi_5\rangle
={\cal D}_{22}= {\cal D}^{-1}_{11}|{\cal D}^{-1}|^{-1},
\eeq
with $det({\cal D}^{-1})\equiv |{\cal D}^{-1}|$.
Recalling the Dzyaloshinskii relation \cite{abr},
\beq
{\cal D}(\omega_n)=D^R(i\omega_n),
\eeq
with the retarded (real-time) Green's function $D^R$, we find 
the excitation energy of the kaon modes are given by $D^{R-1}_{ij}=0$ or 
$det({\cal D}^{-1})=0$.

Similarly 
\beq
\langle \phi_4\phi_5\rangle={\cal D}_{12}=
-{\cal D}_{21}^{-1}|{\cal D}^{-1}|^{-1},\quad {\rm and} \quad
\langle \phi_5\phi_4\rangle={\cal D}_{21}=
-{\cal D}_{12}^{-1}|{\cal D}^{-1}|^{-1}.
\eeq

\subsection{Calculation of $n_K^0,n_K^1$ ,$n_K^2$}

Consider the following quantities;
$n_K^1\equiv \langle K^+K^-\rangle$, 
$n_K^0\equiv \langle (K^+\dot K^--\dot K^+K^-)\rangle$ and 
$n_K^2=n_K^0+2\tilde\mu n_K^1$. Hence it is sufficient to evaluate
$n_K^0$ and $n_K^1$. By way of the kaon Green's functions they are
represented as
\begin{eqnarray}
n_K^1= \langle K^+K^-\rangle&=&
1/2\beta^{-1}\sum_n\int\frac{d^3p}{(2\pi)^3}({\cal D}_{11}+{\cal D}_{22})
\nonumber\\
&=&1/2\beta^{-1}\sum_n\int\frac{d^3p}{(2\pi)^3}[|{\cal D}^{-1}|^{-1}
({\cal D}_{11}^{-1}+{\cal D}_{22}^{-1})],
\end{eqnarray}
and
\beq
n_K^0=\langle (K^+\dot K^--\dot K^+K^-)\rangle
=\beta^{-1}\sum_n\int\frac{d^3p}{(2\pi)^3}2\omega_n
|{\cal D}^{-1}|^{-1}{\cal D}^{-1}_{12}.
\eeq
Generally it is rather complicated to evaluate these quantities, but
we shall see that we can easily do them under the approximation
$c_3=0$. In this case, ${\cal D}_{11}^{-1}={\cal D}_{22}^{-1}$, thereby
the determinant $|{\cal D}^{-1}|$ is
factorized into 
$({\cal D}_{11}^{-1}+i{\cal D}_{12}^{-1})
({\cal D}_{11}^{-1}-i{\cal D}_{12}^{-1})$. Hence we find
\begin{eqnarray}
det({\cal D}^{-1})&\rightarrow& 
\left(\frac{1}{{\cal D}_{11}^{-1}+i{\cal D}_{12}^{-1}}+
\frac{1}{{\cal D}_{11}^{-1}-i{\cal D}_{12}^{-1}}\right)
(2{\cal D}_{11}^{-1})^{-1}\nonumber\\
&=&\left(\frac{1}{{\cal D}_{11}^{-1}+i{\cal D}_{12}^{-1}}-
\frac{1}{{\cal D}_{11}^{-1}-i{\cal D}_{12}^{-1}}\right)
(-2i{\cal D}_{12}^{-1})^{-1}.
\end{eqnarray}
Note that 
${\cal D}_{11}^{-1}\pm i{\cal D}_{12}^{-1}
=(\omega_n\pm iE_-)(\omega_n\mp iE_+)$, then $n_K^0$, $n_K^1$ can be
written as
\beq
n_K^1=\beta^{-1}\sum_n\int\frac{d^3p}{(2\pi)^3}
\left[\frac{1}{E_++E_-}
\left(\frac{E_-}{\omega_n^2+E_-^2}+\frac{E_+}{\omega_n^2+E_+^2}\right)\right],
\eeq
and
\beq
n_K^0=\beta^{-1}\sum_n\int\frac{d^3p}{(2\pi)^3}
\left[\frac{4\omega_n^2}{E_++E_-}
\left(\frac{1}{\omega_n^2+E_-^2}-\frac{1}{\omega_n^2+E_+^2}\right)\right].
\eeq
Using the relation
\beq
\sum_{n=-\infty}^{\infty}\frac{1}{x^2+n^2}=\frac{\pi}{x}\coth\pi x,
\eeq
we finally get
\beq
n_K^1=\int\frac{d^3p}{(2\pi)^3}\frac{1}{E_++E_-}
\left[1+\frac{1}{e^{\beta E_-}-1}+\frac{1}{e^{\beta E_+}-1}\right],
\label{nk1}
\eeq
and
\beq
n_K^0=\int\frac{d^3p}{(2\pi)^3}\frac{1}{(E_++E_-)/2}
\left[-(E_--E_+)/2-\frac{E_+}{e^{\beta E_+}-1}+\frac{E_-}{e^{\beta
E_-}-1}\right].
\label{nk0}
\eeq
The first terms in Eqs.~(\ref{nk1}) and (\ref{nk0}) mean the
contributions from the
zero-point fluctuation to be renormalized. 
Finally $n_K^2=n_K^0+2\tilde\mu_Kn_K^1$ can be
written as
\beq
n_K^2=\int\frac{d^3p}{(2\pi)^3}\frac{1}{(E_++E_-)/2}
\left[-\frac{E_+-\tilde\mu_K}{e^{\beta E_+}-1}
+\frac{E_-+\tilde\mu_K}{e^{\beta E_-}-1}\right].
\label{nk2}
\eeq
It is to be noted that there is no contribution by the zero-point
fluctuation in Eq.~(\ref{nk2})
due to the vector nature of the $KN$ interaction.


\end{document}